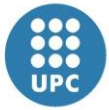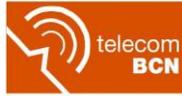

Escola Tècnica Superior d'Enginyeria
de Telecomunicació de Barcelona

UNIVERSITAT POLITÈCNICA DE CATALUNYA

# MASTER'S THESIS

# Design and implementation of an Android application to anonymously analyse locations of the citizens in Barcelona

*Estudis: Master Degree in Telecommunications Engineering*
*Autor:* Ángel Torres Moreira
*Director/a:* Mónica Aguilar Igartua
*Co-Director/a:* Silvia Puglisi

*Date: July 2015*

## Resum del Projecte

L'aplicació "MobilitApp" és capaç d'obtenir les dades de mobilitat i el tipus d'activitat que realitza una persona, en segon pla i mitjançant localitzacions periòdiques de manera síncrona.

El treball principal d'aquest projecte ha estat desenvolupar eines que facilitin l'explotació de la informació obtinguda, afegir elements que facin més atractiu l'ús de l'aplicació i difondre la mateixa a una major audiència.

Amb aquests nous procediments, s'ha aconseguit augmentar la quantitat d'usuaris connectats, millorar la seguretat amb la qual es manegen la informació sensible de l'usuari i establir els canals, que en treballs futurs, serviran per afegir funcionalitats addicionals a l'aplicació.

## Resumen del Proyecto

La aplicación MobilitApp es capaz de obtener los datos de movilidad  y el tipo de actividad que realiza una persona, en segundo plano y mediante localizaciones periódicas de manera síncrona.

El trabajo principal de este proyecto ha sido desarrollar herramientas que faciliten la explotación de la información obtenida, añadir elementos que hagan más atractivo el uso de la aplicación y difundir la misma a una mayor audiencia.

Con estos nuevos procedimientos, se ha conseguido aumentar la cantidad de usuarios conectados, mejorar la seguridad con la que se manejan la información sensible del usuario y establecer los canales, que en trabajos futuros servirán, para añadir funcionalidades adicionales a la aplicación.

# Abstract

The MobilitApp application is able to obtain mobility data and type of activity performed by a person. It runs in the background and stores the information in synchronous periodic locations.

The main work of this project was to develop tools that facilitate the exploitation of the information obtained, add elements that make it attractive to use the application and spread it to a wider audience.

With these new procedures, we manage to increase the number of connected users, improve security with which sensitive user information is managed and establish channels that will be used to add additional functionality to the application in future works.

# Table of Contents

# List of figures

# List of tables

# Glossary of Acronyms

**API** Application Programming Interface

**SDK** Software Development Kit

**APK** Android Application Package

**APP** Application

**ATM** Autoritat del Transport Metropolità

**AMB** Àrea Metropolitana de Barcelona

**GUI** Graphical User Interface

**GPS** Global Positioning System

**JSON** JavaScript Object Notation

**UPC** Universitat Politècnica de Catalunya

**Wi-Fi** Wireless Fidelity

**WPS** Wi-Fi-based Positioning System

**XML** eXtensible Markup Language

**DGT** Dirección General de Tráfico

**GCM** Google Cloud Messaging

**DNS** Domain Name Service

# Chapter 1
# Introduction

# 1. Introduction

In this project, we are going to develop an Android application to obtain data regarding the mobility of citizens in Barcelona. In collaboration with ATM, this project's main objective is to use this data to determine mobility patterns that could be used to improve current transportation infrastructure.

The Smart City concept was actively used for developing this project as we have taken advantage of the Barcelona Open Data service, which serves real-time information of some services in Barcelona.

This application will store confidential information with the locations of the users on a server, which will be focused on the exploitation of the obtained data.

Android Platform is used to develop MobilitApp for two main reasons:

- **-** Large Audience: According to [1] in Spain at December 2014, 83% of the devices use Android.

- **-** Google APIs: Google provides developers large amount of APIs (Application Programming Interface) to add different features to Apps. MobilitApp uses: Maps, Places, Directions, Location and Activity Recognition.

This study is part of the larger EMRISCO project and will help ATM (Autoritat del Transport Metropolità) providing mobility data to improve the current transportation infrastructure.

After the previous work [2], where the detection algorithms were improved and more features were added, the need of having tools to exploit this information arises. This project deploys these tools which will focus on:

- Exploitation of the data obtained.
- Improve the user's experience.
- Establish secure methods to send and store sensible information.
- Spreading the application to a wider audience.

In order to achieve this goals, we incorporate new features to the MobilitApp such as a relational database in a private server, RSA encryption when uploading the information, the publication of the app to the best know repository: Google Play Store and many other features which will be explained in later chapters.

This document is structured with seven chapters plus references and annexes.

The first sections will explain some Android fundamentals such as fragments, services, the Google APIs used, and then we will make a summary of the work previously performed.

In the next chapters, we analyse the features added such as the server, what are its properties and its functions in the developing of the project.

Later, we describe the security method used to transfer information to the server. To complete the objectives, we describe some of the tools that where developed to enhance and spread the use of the application such as to show in the map the real time traffic state and traffic incidences or put the mobile in silence mode when the application detects that the user is in a vehicle.

We present the tools used to spread the use of the application: the design of a web page, the implementation of Google Cloud Messaging to establish a communication channel between the server and all the users of the app, and the publication of the app to the official Android Store.

In the annexes we include a brief guide of how to install the server and we show the code of the web pages designed.

MobilitApp final degree project is part of a National Spanish project [3] and can be found in: ---.

# Chapter 2
# The MobilitApp Application

# 2. MobilitApp

This sections gives a brief introduction to the main android concepts necessaries to know also describes the main features implemented in previous works that will serve as a starting point of this project.

## 2.1. Android general concepts

### 2.1.1. Activity

An activity [4] is an application component providing a view users can interact with. An application consists of multiple activities. The "main activity" is presented to the user when the application is launched for the first time. In our case, MobilitApp chooses between two "main" activities: Login Activity and the Main Activity. The first one is launched the first time MobilitApp is opened and the second one the following times. This Main Activity will start Location service and Activity Recognition service.

Each activity can start another activity in order to perform different actions. Each time a new activity starts, the previous activity is stopped, but the system preserves the activity in a stack (i.e. the "back stack"). When a new activity starts, it is pushed onto the back stack and takes user focus. When the user is done with the current activity and presses the Back button, it is popped from the stack and destroyed and the previous activity resumes.

Activities must be declared in the manifest file in order to be accessible to the system. This manifest file presents essential information about the app to the Android system. Information like: minimum SDK (Software Development Kit) version, permissions, activities, services...

#### 2.1.1.1. Lifecycle

It is essential to manage the lifecycle of our activities in order to develop a strong and flexible application. An activity has three states:

- **Resumed:** The activity is in the foreground of the screen and has user focus.

- **Paused:** Another activity is in the foreground (visible) and has focus, but this one is still visible. Paused activities are completely alive but can be killed by the system in low memory situations.

- **Stopped:** The activity is running in the background and is no longer visible by the user. Stopped activities are also alive and can be killed by the system.

In order to manage the lifecycle of our activity, we need to implement the callback methods. Callback methods can be overridden to do the appropriate work when the state of the activity changes. The most important callback methods are:

```java
public class ExampleActivity extends Activity {
    @Override
    public void onCreate(Bundle savedInstanceState) {
        super.onCreate(savedInstanceState);
        // The activity is being created.
    }
    @Override
    protected void onResume() {
        super.onResume();
        // The activity has become visible (it is now "resumed").
    }
    @Override
    protected void onPause() {
        super.onPause();
        // Another activity is taking focus (this activity is about to be "paused").
    }
    @Override
    protected void onDestroy() {
        super.onDestroy();
        // The activity is about to be destroyed.
    }
}
```

*onCreate()* method is called when the activity is created for the first time. This is where all the set up needs to be done such as create views, bind data to lists, and so on.

*onResume()* method is called just before the activity starts interacting with the user.

*onPause()* method is called when the system is about to start resuming another activity.

*onDestroy()* method is called before the activity is destroyed.

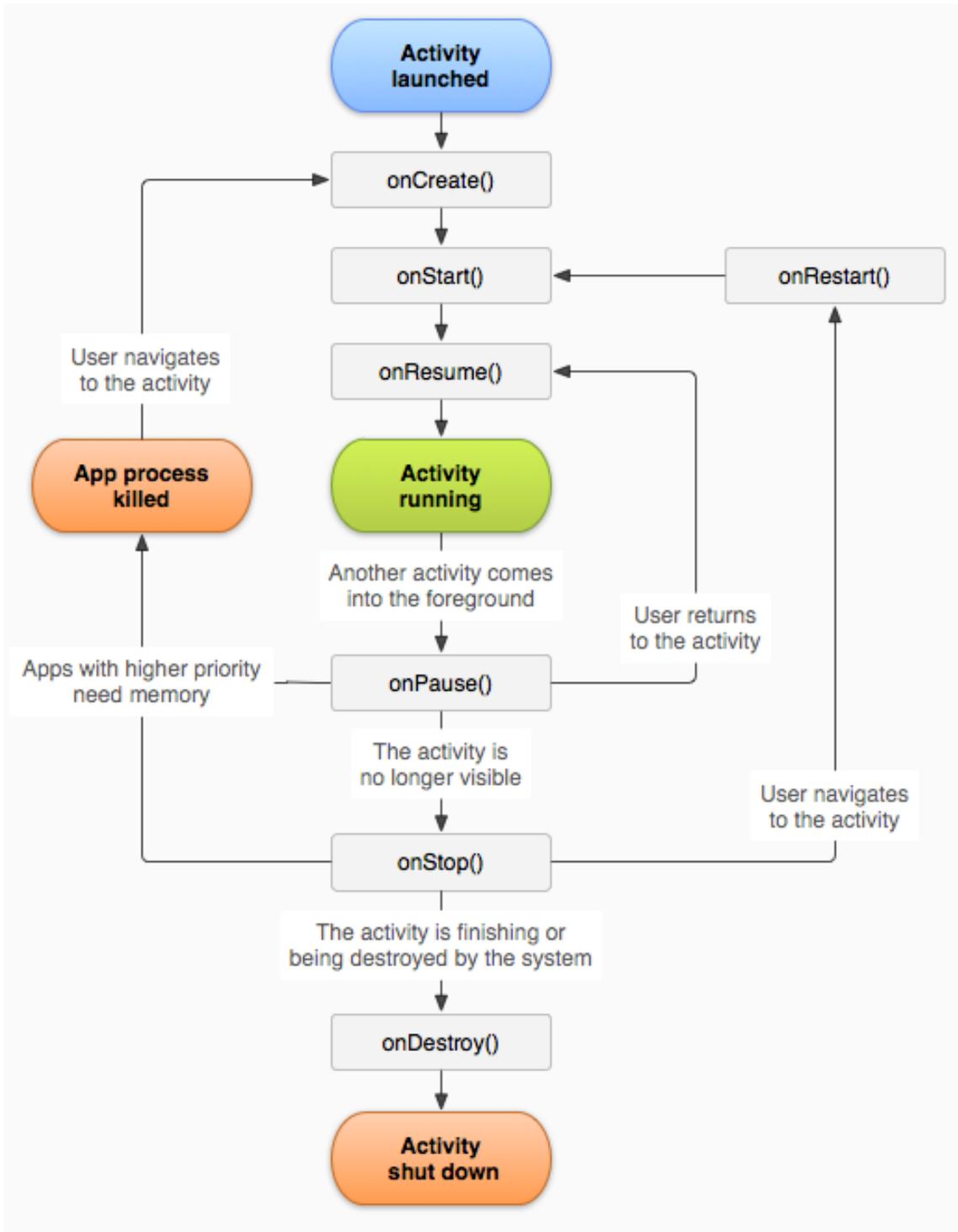

### 2.1.2. Fragment

A Fragment [5] represents a portion of the user interface in an Activity. Multiple fragments can be combined in a single activity. Each fragment has its own lifecycle and can be added or removed while the activity is running.
MobilitApp uses fragments in all the activities because allows more dynamic and flexible designs.
There are two ways we can add a fragment to the activity layout:

- **Declaring the fragment inside the activity's layout file**

In this case, we can specify layout properties for the fragment as if it were a view. This is how we add the map on the activity.

```
<fragment
      android:id="@+id/map"
      android:layout_width="match_parent"
      android:layout_height="match_parent"
      class="com.google.android.gms.maps.MapFragment" />
```

- **Programmatically add the fragment to an existing ViewGroup**

At any time the activity is running, we can add fragments to our activity layout. We simply need to specify a ViewGroup in which to place the fragment.
Adding, replacing or removing actions are called fragments transactions and we need an instance of *FragmentTransaction* [6] to use methods such as: add, remove or replace.

```
getSupportFragmentManager().beginTransaction().add(R.id.containerhist, new
HistoryFragment()).commit();
```

In this case, the *ViewGroup* is a *FrameLayout* declared in the activity's layout file and we use the "id" as a reference.

```
<FrameLayout
      android:id="@+id/containerhist"
      android:layout_width="match_parent"
      android:layout_height="wrap_content"
      android:layout_below="@id/toolbar"
      android:visibility="gone" >
</FrameLayout>
```

The fragment can access to its activity instance (and use activity methods) using *getActivity()*

### 2.1.2.1. Lifecycle

The most important callback methods are the same as with Activity plus one more:

*onCreateView()* method is called when it is time for the fragment to draw its user interface for the first time.

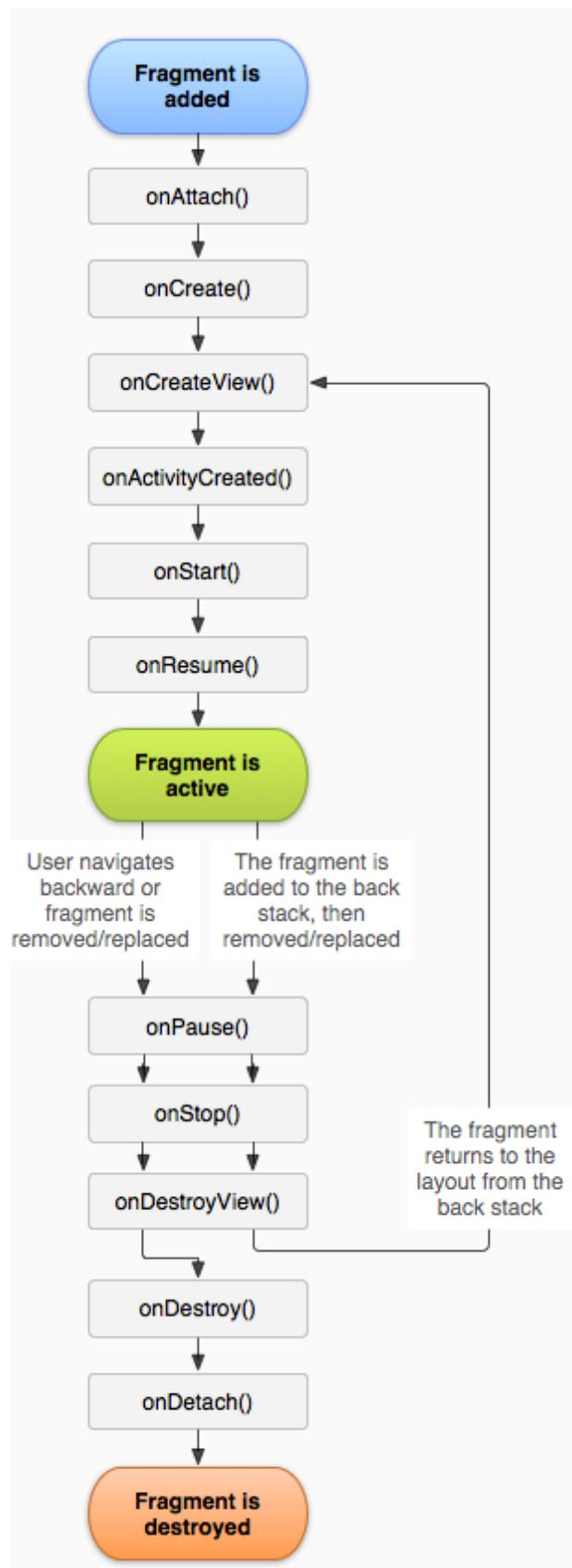

Figure 2-2: Fragment Lifecycle

### 2.1.3. Service

A Service [7]is an application component that can perform long-running operations in the background and does not provide a user interface. A service can take two forms:

- **Started:** A service is started when an activity or a fragment starts it by calling *startService()* method. Once started, a service can run in the background indefinitely, even if the activity or fragment is destroyed. This is the form used by MobilitApp for Location Service and Activity Recognition Service.

- **Bound:** A service is "bound" when an activity or a fragment binds to it by calling *bindService()* method. A bound service offers a client-server interface that allows activities/fragments to interact with the service. This form is not used by MobilitApp.

There are two types of service: the normal (i.e. *Service*) and the *IntentService*.

The main difference between the two is that *Service* uses the application's main thread so it could slow the performance of any running activity. IntentService uses a worker thread to handle all start requests, one at a time. This work queue passes on intent at a time to our *onHandleIntent()* callback method.

#### 2.1.3.1. Lifecycle

The most important callback methods are:

*onStartCommand()* method is called every time the service is started by calling *startService()*

*onHandleIntent()* method is called to process intents in the worker thread. When all requests have been handled stops itself.

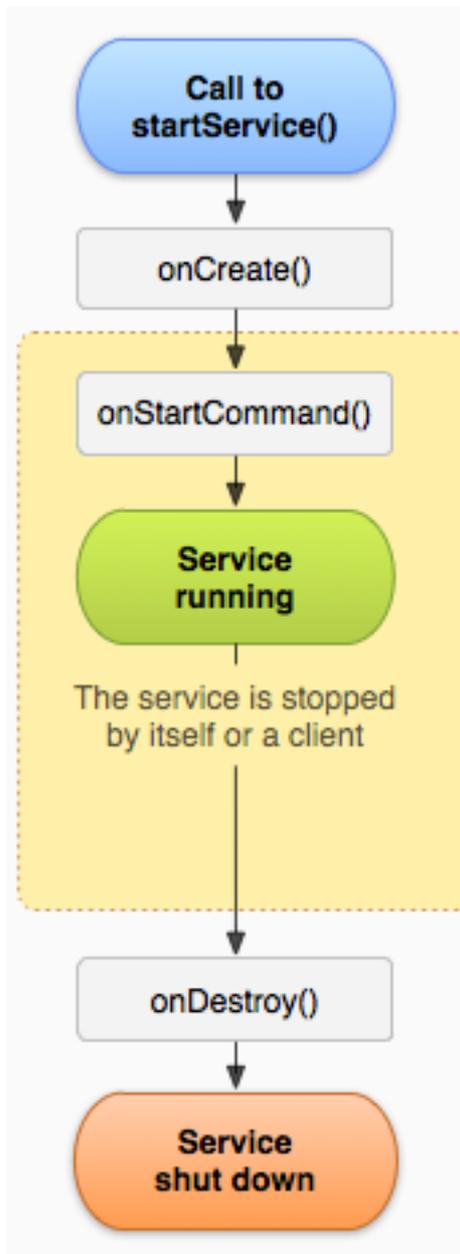

Figure 2-3: Service Lifecycle

### 2.1.4. Location Sources in Android devices

The selection of location sources is directly related to meet, on the one side, the low power consumption criterion, and on the other side, the specified requirements for mobility data. There are at least four location sources available in almost every Android device: GPS, WPS (Wi-Fi-based Positioning System), Cell ID and sensors. Each one of these sources has its own features regarding average power consumption, accuracy and coverage which basically depends on how location data is obtained as well as the minimum optimal update intervals to achieve with an acceptable level in quality of data.

Unfortunately, Android platform is not well documented regarding detailed profiles of average power consumption of each one of the location sources. For that reason, we are going to use approximations of their power usage profiles in order to select those that could meet our low power consumption criterion.

#### 2.1.4.1. Global Positioning System

GPS consists of up to 24 or more satellites broadcasting radio signals providing their locations, status and timestamp. The GPS receiver in the Android device can calculate the time difference between broadcast time and the time radio signal is received. When the device knows its distance from at least four different satellite signals, it can calculate its geographical position.

- **Average Power consumption:** GPS receiver works as an independent-powered component in the Android platform with the unique objective to obtain location samples. Therefore, all the battery power consumed by GPS is consequence of a user positioning operation. According to several references, we can estimate that the average power consumption of an active GPS with a location update interval between 5 and 15 seconds is approximately 125 to 145 mA.

- **Coverage:** GPS location coverage is probably the most limited of all four available sources. Although is a source that cover the whole globe, it only provides good performance in outdoor scenarios, because it requires satellites' visibility. GPS does not work in indoor locations (may provide some location data in very few of them if it has some kind of visibility of satellites, but with very poor performance) like buildings or undergrounds. Even in outdoor environments, GPS may suffer from poor performance, especially in cities, because objects like buildings, trees and other obstacles can overshadow visible satellites, decreasing its performance. Therefore, GPS range of coverage can be quite limited in a metropolitan region, compared to other sources, specifically if we want a continuous position tracking of the citizens.

- **Accuracy:** In outdoor scenarios with good satellite visibility, GPS is the positioning system that can provide the most accurate location of all four existing sources. In perfect conditions GPS can achieve a precision higher than 3 meters, but in optimal conditions the range is between 3 and 10 meters. In non-optimal conditions, i.e. lower visibility on the satellites, GPS accuracy range is typically between 30 and 100 meters. GPS is also capable to provide an estimation of altitude and speed along with location.

### *2.1.4.2.    Wi-Fi-based Positioning System*

WPS sends a location request to Google location server with a list of MAC that are currently visible by the device, then the server compares this list with a list of known MAC addresses of the device itself, and identifies associated geocoded locations. After that, Google server uses these locations to triangulate the approximate location of the user.

-   **Average Power consumption:** Wi-Fi receiver works as an independent-powered component in the Android platform but multiple-purpose that is not restricted only to obtain location samples, as happens with GPS receiver. Average power consumption of WPS represents only a small part of power usage of Wi-Fi services. According to multiple references, we are going to estimate that the average power consumption of an active WPS with a location update interval around 20 seconds is approximately 25 mA.

-   **Coverage:** WPS location coverage is less broad than GPS but is more versatile. Its coverage it is not global but block level, because it has to be in the range of a wireless signal to be able to provide location updates. However, within this range, it can provide good performance both outdoor and indoor scenarios. Therefore, WPS range of coverage can be very useful when tracking continuous locations within cities and towns.

-   **Accuracy:** In indoor scenarios, WPS is the positioning system that can provide the most accurate location of all four existing sources. In normal conditions, WPS can achieve a precision between 3 and 10 meters. In outdoors, the accuracy of GPS decreases significantly being able to provide a range between 25 and 150 meters, depending on the external conditions of the environment and the signal strength.

### *2.1.4.3.    Cell-ID based localization*

Cell-ID based localization uses both the Location Area Code and Cell-ID that the Base Transceiver Station broadcasts. Android devices are always receiving these broadcast messages, which means that they always know their Cell-ID at any time. Knowing this, it can obtain an approximation of its actual location using the geographical coordinates of the corresponding Base Station.

-   **Average Power consumption:** Since Android devices always know their Cell-ID; the average power consumption consequence of this positioning system is extremely low.

-   **Coverage:** Cell-ID location coverage is global, as happens with GPS, and versatile, as happens with WPS, which implies that has the best coverage of all four existing location sources. Therefore, cell-ID has an adequate coverage range to continuously track location of citizens in the metropolitan region.

-   **Accuracy:** Cell-ID based location system accuracy depends on cells size in the network that are connected with user's Android device. This means that in those regions with smaller cells (typically in cities and metropolitan regions), cell-ID location will provide good performance achieving an accuracy range between 50 and 200 meters. Moreover, in regions with bigger cells (typically in rural areas) performance will significantly decrease, and cell-ID will only achieve an accuracy range between 1000 meters and several kilometres.

### 2.1.4.4.    Sensors

- **Average Power consumption:** Android devices have usually several built-in sensors. However, not all of them can be used for positioning processes, being only accelerometer, magnetic field, orientation, and gyroscope that one that can be used for such purpose. Based on their specifications, the power consumption of these sensors are the following:

    o **Accelerometer** - 0.23 mA
    o **Magnetic Field** - 6.8 mA
    o **Orientation** - 13.13 mA
    o **Gyroscope** - 6.1 mA

  If all sensors were used together at the same time to obtain locations updates, we would have an average power consumption of 26.26 mA.

- **Coverage:** Sensors have a wide coverage because they can process multiple different types of data. From acceleration, that covers an individual range, to magnetic field, that covers a global range, sensors can provide data to improve location samples in almost any scenario.

- **Accuracy:** It depends on the type of sensor and how is used, which means it is infeasible to provide real accuracy values. However, according to Google, sensors improve the accuracy obtained with just simple WPS and Cell-ID.

### 2.1.4.5.    Android Location Tools

#### Fused Location Provider

This API considers all hardware components as single location sources layer that can be used according to predetermined criteria implemented on it. It offers four main features:

- **Simple API:** We can specify high-level needs like "high accuracy" or "low power", instead of having to worry about location providers.
- **Immediately available:** Gives immediate access to the best and most recent location.
- **Power-efficiency:** Minimizes app's use of power. Based on all incoming location requests and available sensors, fused location provider chooses the most efficient way to meet those needs.
- **Versatility:** Meets a wide range of needs, from foreground uses that need highly accurate location to background uses that need periodic location updates with negligible power impact.

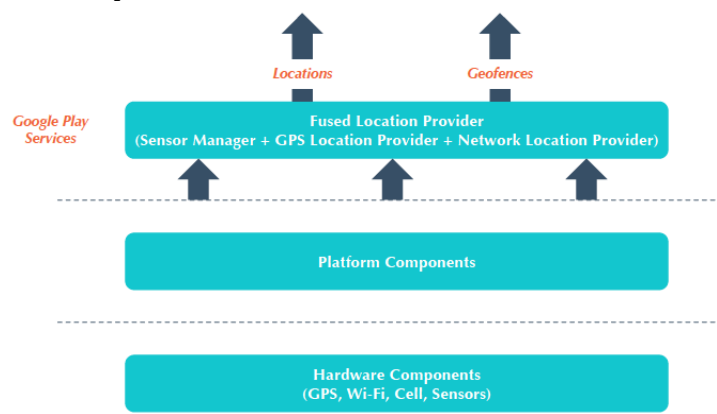

**Figure 2-4: Fused Location Provider**

**Location Request**

Location Request [8] is a data object that contains quality of service parameters for requests to the Fused Google Provider. The main parameters are:

- **Desired interval time:** This parameter establishes the ideal interval time between location updates. However, it is not an exact interval, which means that locations can be obtained in lower or higher interval times, depending on external conditions. According to our design criteria we should set this parameter to 20 seconds.

- **Fastest interval time:** This parameter establishes the minimum interval time between location updates. Unlike desired interval time, this is an exact interval, which means that location updates will not be obtained in lower intervals of time than set by this parameter. According to our design criteria we should also set this parameter to 20 seconds.

- **Priority mode:** This parameter establishes the priority of the location requests based on the location sources that we want to use and the accuracy of location samples we require. There are the following modes:

    o **High Accuracy:** This is the highest priority mode. Using this mode Location Client will obtain the best location available using all the location sources available together. This is also the mode with higher average power consumption because it is the only one that actually uses the GPS.

    o **Balanced Power Accuracy:** This mode has lower priority than High Accuracy mode. This mode only uses WPS, cell-ID and sensors to obtain location updates with a "block" level accuracy while keeping average power consumption very low.

    o **Low Power:** This mode has a lower priority than Balanced Power Accuracy mode. This mode only uses WPS, cell-ID and sensors to obtain location updates with a "city" level accuracy (10km) while keeping average power consumption extremely low.

    o **No Power:** This is the lowest priority mode. This mode does not use any location source to obtain location updates. Instead, obtains location updates requested by other third party applications in the Android platform. In this way the average power consumption is the lowest possible.

| PRIORITY | INTERVAL (s) | BATTERY DRAIN PER HOUR (%) | ACCURACY (m) |
|----------|--------------|----------------------------|--------------|
| HIGH_ACCURACY | 5 | 7.25 | 10 |
| BALANCED_POWER_ACCURACY | 20 | 0.6 | 100 |
| LOW_POWER | - | small | 10.000 |
| NO_POWER | N/A | small | variable |

Table 2-1: Location Request priority modes

### 2.1.5. Languages

It is always a good practice to present the application with more languages; thereby we can reach more users in Barcelona. MobilitApp [9] is available in three languages (i.e. Spanish, Catalan and English) depending on the language their mobile phone is set. It cannot be changed, so if the user's phone is in Catalan, MobilitApp dialog and text will be in Catalan too. If user's language is neither Spanish nor Catalan, the default language is set (i.e. English).

To make this possible we need to create locale directories and string files. Within *res/* directory are subdirectories for various resource types (e.g. Layout resources are saved in *res/layout,* Drawable resources are saved in *res/drawable...)*. String Resources define strings and string arrays and are saved in *res/values.* This is the default directory, so if we don't have any additional *values* directory this will be the one Android would use. An example of a bad practice is the following: all the strings in *values* directory are written in Spanish but user's mobile locale are set in English. In this situation user does not understand anything because MobilitApp dialogs and text are in Spanish and user only understands English. To solve this, we need to create additional *values* directories.

These additional *values* directories are followed by a hyphen and the ISO language code. If the language is spoken in different countries we add another hyphen and the region (preceded by lowercase "r".

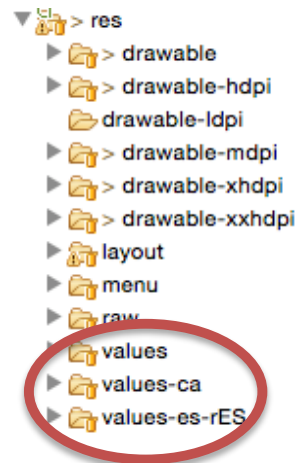

- **Spanish *values* directory:** values-es-rES
- **Catalan *values* directory:** values-ca

**Figure 2-5 : Available Languages**

Now we can create a *strings.xml* file in each directory with its respective language and then add string values for each locale into the appropriate file.
At runtime, the Android system uses the appropriate set of string resources based on the locale currently set for the user's device.

For example, the following is a string resource to be defined in the three languages.

```xml
<string name="acep">Aceptar</string>
<string name="cancel">Cancelar</string>
```

Finally, we can reference our string resources in the code using the resource name defined by the string element's *name* attribute.

For example:

When we only need to supply the string resource to a method that requires a string.

```
builder.setNegativeButton(R.string.cancel, null);
```

Or, when we need the string itself (i.e. the text) and not the resource.

```
value[0] = getActivity().getResources().getString(R.string.calorie);
```

### 2.1.6. Shared Preferences

Android provides several options to save persistent application data [10]. Choosing between these options depends on our specific needs, such as whether the data should be private or accessible to other applications and how much space our data requires. Android provides five options to store data:

- **Shared Preferences:** Store private primitive data in key-value pairs.
- **Internal Storage:** Store private data on the device memory.
- **External Storage:** Store public data on the shared external storage.
- **SQLite Databases:** Store structured data in a private database.
- **Network Connection:** Store data on the web with your own network server.

MobilitApp use three of these five options:

- **Shared Preferences:** Store simple data such as profile information.
- **External Storage:** Store JSON (JavaScript Object Notation) files in *Downloads/* device directory.
- **Network Connection:** Store JSON files using Google Cloud Storage.

External Storage is explained in Segment section and Network Connection in Upload Section. So now, we need to explain the other store option used by MobilitApp: Shared Preferences.

The *SharedPreferences* class [11]provides a general framework that allows us to save and retrieve persistent key-value pairs of primitive data types. We can use *SharedPreferences* to save any primitive data: Booleans, floats, int, longs and strings. This data will persist across user sessions, even if the application is closed or killed.

To get a *SharedPreferences* object for our applications we use the following method with two parameters: the name (i.e. filename) and the operating mode by default (i.e. MODE_PRIVATE).

```
prefsGoogle = getSharedPreferences("google_login", Context.MODE_PRIVATE);
```

If the files does not exist it will be created with the filename indicated.
In Private Mode the created file can only be accessed by the calling application (i.e. MobilitApp).

- Writing or Modification Values

1. Call *edit()* to get a *SharedPreferences.Editor* [10].

```
SharedPreferences.Editor editorG = prefsGoogle.edit();
```

2. Add values with methods such as *putBoolean()* or *putString().* The first parameter is the key and the second is the value.

```
editorG.putString("birthday", p.getBirthday());
```

3. Commit the new values with *commit().*

```
editorG.commit();
```

- Reading Values

1. We only need to use methods such as *getBoolean()* or *getString().* The first parameter is the key and the second is the default value that will be returned when the key does not exist.

```
prefsGoogle.getString("nombre", "User");
```

### 2.1.6.1. Output File Example

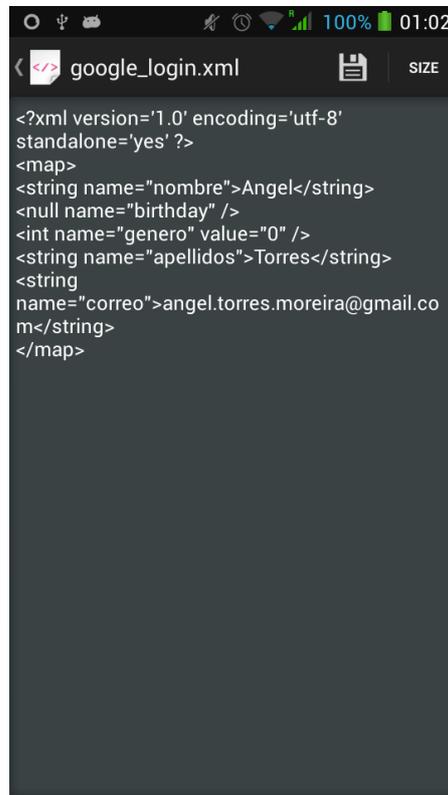

Figure 2-6: Shared Preferences file example

## 2.2.  MobilitApp fundamental concepts.

This section describes the main concepts that should to be appreciated from previous work done in the application. We discuss how the application identifies the user activity, defining the tools that the application uses.
Next, we detail the format used to store the information in the terminal.

### 2.2.1.  Activity Recognition

One of the main achievements in [2]was the ability to recognize the activities in which the users move. MobilitApp uses Google APIs with user's geo positions as a first approach for the detection of the activity.

The MobilitApp uses Google Activity Recognition API. This API provides a low power mechanism that can provide periodic updates of detected activity types by using low power sensors (basically accelerometer). We are going to process these samples in order to increase accuracy and efficiency. The most important classes that we are going to use from this API are the following:

-  Activity Recognition Result

This class [12] returns a list with all probable activities. It also provides a method that directly returns the most probable activity along with its confidence level called *getMostProbableActivity().*

-  Detected Activity

This class [13] returns an integer that indicates which activity type has been detected by Activity Recognition API.

The possible detected activities by this API are:

-  *on_foot:* Activity type returned if the citizen is either walking or running.

-  *bicycle:* Activity type returned if the citizen is on a bicycle.

-  vehicle: Activity type returned if the citizen is on a motor vehicle (e.g. car, motorbike, bus,…).

-  still: Activity type returned if the citizen is not moving.x

-  unknown: Activity type returned if Activity Recognition API is not capable to estimate the actual activity.

As a first approach, the activity detection algorithm makes a temporal sampling of the results of the API.
Every 20 seconds consults the results of the Activity Recognition API then after 2 minutes, checks which is the most repeated result in the last samples.

In order to get a more accurate recognition besides other kind of activities such as metro, bus, tram and train, more algorithms were developed. This algorithms merge the results of Google Activity Recognition API with queries to other APIs and Open Data services. These algorithms takes into account a few variables appearing when one works with geo-locations such as:

-    Accuracy of the GPS:
When a terminal is underground, the GPS accuracy decreases (error margin increases) So that one can identify if the user is in the metro by checking periodically this value.

-    Location of POI (Points of interest):
An example of POI that helps the algorithms are the position of bus stops. With this information, the algorithms predict if the user goes in a bus, tram, and train.

-    Directions:
Having identified the position of the stops, the algorithms use Google Directions API to check if there is a route between two points given by the terminal's location. This information will help to the accuracy of the activity recognition algorithms.

### 2.2.2. Segments

A segment can be defined as a group of consecutive user's locations where its activity remains unchanged. The main components of a segment are the locations described as a pair of latitude/longitude. These pairs will mark the path that the segment follows.
An example of segment is represented below:

```
{
"segments": [
  {
   "activity": "on_foot",
   "distance (m)": 49.69776445992602,
   "duration (s)": 142,
   "speed (Km\/h)": 1.2599432,
   "first time": "09:46:44",
   "last time": "09:49:07",
   "location": [
    41.441145,
    2.1659081,
    "09:46:44",
    41.4410568,
    2.1660705,
    "09:47:11",
    41.441012,
    2.1661082,
    "09:47:32",
    41.4409738,
    2.1661926,
    "09:48:13",
    41.440959,
    2.1662142,
    "09:48:34",
    41.4410113,
    2.1663986,
    "09:49:07"
   ]
  },

  (...)

 ]
}
```

Figure 2-7: : Format of a Segment

The user's segments are stored in a local folder first before processing and sending the information to the server. These files will be the base for other parts of the application for instance "History" where the users can draw on the Map their locations classified by day and detected activities.

This format of the segment was slightly modified in this work. Besides getting the location latitude, longitude and time, we save also the GSM and Wi-Fi power measured when the location is saved. An example of the new format is shown in Figure 2-8: New Segment

```
{
  "duration (s)": 560,
  "speed (Km\/h)": 2.9486754846691303,
  "last time": "10:27:53",
  "first time": "10:19:10",
  "location": [
    41.4007067,
    2.1629655,
    "10:19:10",
    "-105\/-55",
    41.4007211,
    2.1629995,
    "10:19:36",
    "-85\/-200",
    41.4004902,
    2.1633019,
    "10:20:07",
    "-85\/-200",
    41.4004788,
    2.1633106,
    "10:20:39",
    "-85\/-200",
    41.4000006,
    2.1636302,
    "10:21:28",
    "-85\/-200",
    41.4001544,
    2.1641032,
    "10:21:59",
    "-85\/-200",
    41.2006757
```

Figure 2-8: New Segment

**Chapter 3**
**The server to store mobility data.**

## 3.  Server

This chapter talks about the main characteristics of the server implemented, which is proposed as an alternative for the storage tool (Google Cloud Storage) used for the previous application development.

One of the main reasons for this new storage proposal is decreasing the budget cost for an implementation in a city with a considerable number of users, where the number of transactions could increase the costs, because of the tariffs in the google service.

A centralized server can provide a fully operational database administration (without cost per operation), which open the possibility to extract valuable information of the user's mobility.

### 3.1.  Server Setup

The equipment used to the development of this paper is a raspberry Pi [14], its configuration is described below. Besides the server, the router providing internet connection should be configured appropriately in order to redirect the traffic to the server.

#### 3.1.1.  Raspberry Pi computer card-sized

Raspberry Pi is the name of an electronic layout that gathers in one credit-card size, the minimal requirements for a computer in a single board configuration. Model B+ is used for the project.

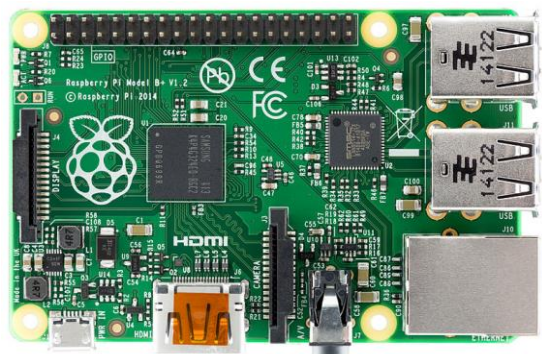

Figure 3-1: Raspberry Pi Model B+

It is based in a Broadcom BCM2835 system on a chip (SoC) including an ARM1176JZF-S 700 MHz processor and 512 MB of RAM.

Raspberry Foundation provides Debian and Arch Linux ARM distributions for the administration of the equipment. The distribution used for the project was "Raspbian".

The main advantage of this equipment is its low power consumption which implies less money to pay, on the other side, the processing power is not comparable to a common dedicated server due to its limited CPU speed and RAM memory. In the performance analysis section, this features will be analyzed to study the feasibility of this hardware.

### 3.1.2. Setup

The Raspberry Pi only includes the hardware, so an operative system is necessary to manage it, all the updates and the network configuration should be properly configured. In the Annex section, step by step manuals are provided.

### 3.1.3. LAMP Server

LAMP [15] refers to **L**inux, **A**pache, **M**ySQL [16], **P**HP-**P**erl-**P**ython, these are the tools necessaries for having a fully operational web server with a relational database support on a Linux system.

This project will use Linux as the operative system (Raspbian distribution [17]), which will manage the hardware resources; Apache which will provide the web-server application; MySQL as the database manager; and PHP as the interpreter to query the database (insert, update, select, delete).
The installation steps on the Raspberry Pi are given in the annex section.

## 3.2. Database

This project uses MySQL as the database engine. In this section the database structure: tables, fields and relations is explained.

### 3.2.1. Tables

- tbl_usuarios

  In this table, all the user information is saved. According to the application premises, the user can save its mobility information anonymously which is why the fields "usu_nombre" where the name of the user is saved can be left empty.
  "usu_hash" is the field where the hash generated when installing the application is kept. This is the main field used to identify one user.
  If the user doesn't login with Google or Facebook, the fields will be filled with the following information:

| Field Name | Content |
|---|---|
| usu_id | AUTONUMERIC |
| usu_nombre | not_set |
| usu_apellido | not_set |
| usu_peso | 0.00 |
| usu_nacimiento | 1900-01-01 |
| usu_genero | not_s |
| usu_mail | notset@notset.com |

There are two more fields in this table: usu_hash y usu_regid. These fields are the mandatory to create a new user.

| Field Name | Content |
|---|---|
| usu_hash | Unique random alphanumeric key created in the user's terminal. This field+ usu_id will differentiate the users |
| usu_regid | Unique key created by the Google Cloud Messaging (GCM) service, this key will identify each user when sending or receiving GCM messages. |

Table 3-2: User's mandatory fields

- tbl_Segmento

  Each segment is associated to a single user. The segments are constructed and processed on the user terminals.

| Field Name | Content |
|---|---|
| seg_id | AUTONUMERIC |
| seg_activity | String to describe the activity of the user in the segment. |
| seg_distance | Distance travelled, in meters. |
| seg_duration | Elapsed time of the segment, in seconds. |
| seg_speed | Mean speed during the segment, in Km/H. |
| seg_firsttime | Timestamp with the first location of the segment. |
| seg_lasttime | Timestamp with the last location of the segment. |
| usu_hash | Unique ID of the associated user. |
| seg_subido | String that identifies if a segment has all its location uploaded correctly. This field will show "OK" if all the locations belonging to this segment are correctly saved. |

Table 3-3: Table "tbl_Segmento"

- tbl_Location

  All the locations given in a pair of Latitude and longitude are stored in this table. As well as segments, the location properties are processed in the user terminals.

| Field Name | Content |
|---|---|
| loc_id | AUTONUMERIC |
| loc_power | String that stores the GSM and WiFi Power of the terminal when the location was gotten. Units are dBm. |
| seg_id | Unique ID of the associated Segment |
| loc_latitude | Decimal field to store the latitude of the location |
| loc_longitude | Decimal field to store the latitude of the location |
| loc_time | Timestamp with the location. |
| loc_date | Date of the location. |

Table 3-4: Table "tbl_Location"

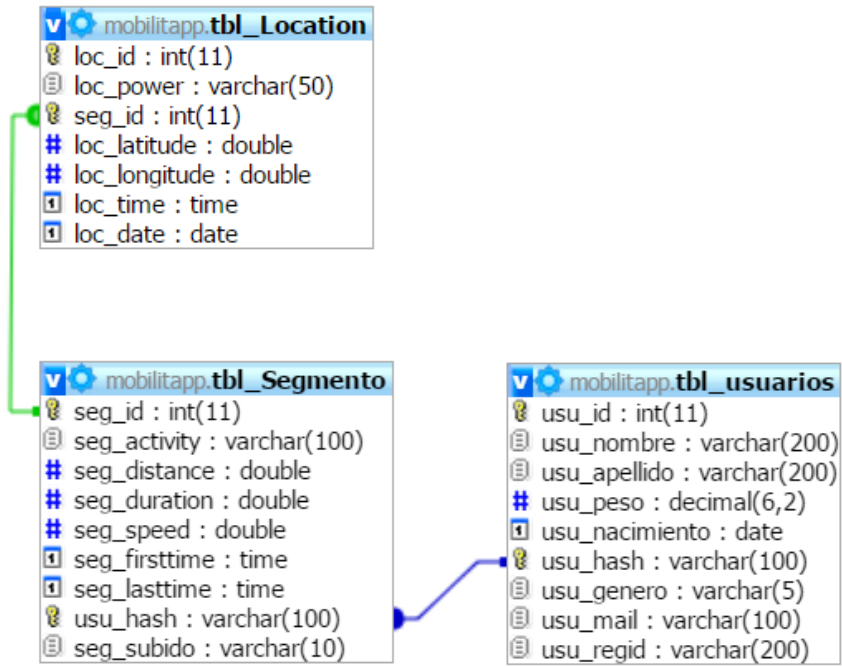

**Figure 3-2: Entity-Relation Model**

**Chapter 4**
**Security and data privacy**

# 4. Security and data privacy

This section will describe the security measures used in the MobilitApp. First presenting the main aspects of the security algorithm (RSA), then explaining the tools used to manage the encryption and decryption operations.

Then, a performance analysis is deployed to decide which the appropriate conditions of the security method are.

## 4.1. The RSA (Reavest-Shamir-Adleman)

The RSA cryptosystem [18] invented by Ron Rivest, Adi Shamir, and Len Adleman, was first publicized in the August 1977 issue of Scientist American. The cryptosystem is most commonly used for providing privacy and ensuring authenticity of digital data. These days RSA is deployed in many commercial systems. It is used by web servers and browsers to secure web track, it is used to ensure privacy and authenticity of Email, it is used to secure remote login sessions, and it is at the heart of electronic credit-card payment systems. In short, RSA is frequently used in applications where security of digital data is a concern.

Public-key cryptography, also known as asymmetric cryptography, uses two different but mathematically linked keys, one public and one private. The public key can be shared with everyone, whereas the private key must be kept secret. In RSA cryptography, both the public and the private keys can encrypt a message; the opposite key from the one used to encrypt a message is used to decrypt it. This attribute is one reason why RSA has become the most widely used asymmetric algorithm: It provides a method of assuring the confidentiality, integrity, authenticity and non-reputability of electronic communications and data storage.

RSA derives its security from the difficulty of factoring large integers that are the product of two large prime numbers. Multiplying these two numbers is easy, but determining the original prime numbers from the total -- factoring -- is considered infeasible due to the time it would take even using today's super computers.

### 4.1.1. Breaking RSA

Full decryption of an RSA ciphertext is thought to be infeasible on the assumption that both of these problems are hard, i.e., no efficient algorithm exists for solving them.

According to [19] , the security of RSA is mostly based in $n$ , the number of bits in the encryption key. In  Table 4-1: Results of the software applied to break RSA. it can be noticed the time and the amount of RAM memory necessary to break the algorithm. The software used to minimize the time to find the factors of arbitrary input integers is called "yafu".

| Bits | Time (s)  | Memory used (MB) |
|------|-----------|------------------|
| 128  | 0.4886    | 0.1              |
| 192  | 3.9979    | 0.5              |
| 256  | 103.1746  | 3                |
| 300  | 1175.7826 | 10.9             |

**Table 4-1: Results of the software applied to break RSA.**

We follow the recommendations made RSA Labs [20]which propose to use at least a 1024 bits length key as it estimates that a $10 million machine, using 2000 computer technology, would take about 3,000,000 years to break a 1024-bit RSA key.

## 4.2. The encryption process

As the MobilitApp uses public key encryption to protect the sensible information of the users when uploading the information to the server, a pair of keys should be created in the server and the public key will be placed in a shared location accessible by the users.
The code below shows the PHP code executed in the server to create private key of 2048 bits and save it with the name "private.pem".

```php
<?php
set_time_limit(300);
$config = array('private_key_bits' => 2048);
$keys = openssl_pkey_new($config);
$priv = openssl_pkey_get_private($keys);
openssl_pkey_export_to_file($priv, 'private.pem');
?>
```

With the private key created and saved in a secure location, the public key is generated executing in a terminal of the server the following command line:

```
$> openssl rsa -in private.pem -pubout -outform DER -out public.der
```

This is the public key that the user will serve to encrypt the messages. Turning back to Android, the application will download the public key, then encrypt the information with it and sent over HTTP requests to the server.
The code below presents an example and the tools needed to perform this operations. All this procedures are part of the Class SaveDatabase which is defined as an asynchronous task:

- Download the public key:

```
downloaFilefromUrl(ruta + "/certs",
"http://mobilitapp.noip.me/certs/public.der", "public.der");
```

The function `downloaFilefromUrl` is called, where "ruta" is the default download directory of the mobile terminal, its definition is described next.

```java
1 public void downloaFilefromUrl(String ruta,String url,String filename){
2        try {
3
4                File folder = new File(ruta);
5                boolean success = true;
6                if (!folder.exists()) {
7                        success = folder.mkdirs();
8                }
9                if (success) {
13                        URL urlPKey = new URL(url);
15                        HttpURLConnection urlConnection = (HttpURLConnection)
urlPKey.openConnection();
16
18                        urlConnection.setRequestMethod("GET");
19                        urlConnection.setDoOutput(true);
20
22                        urlConnection.connect();
23
30                        File file = new File(ruta,filename);
31
33                        FileOutputStream fileOutput = new
FileOutputStream(file);
34
36                        InputStream inputStream =
urlConnection.getInputStream();
37
39                        int totalSize = urlConnection.getContentLength();
41                        int downloadedSize = 0;
42
44                        byte[] buffer = new byte[1024];
45                        int bufferLength = 0;
48                        while ( (bufferLength = inputStream.read(buffer)) > 0
) {
49
50                                fileOutput.write(buffer, 0,
bufferLength);
55                        }
57                        fileOutput.close();
58
59                } else {
61
62                }
64        } catch (MalformedURLException e) {
65                e.printStackTrace();
66        } catch (IOException e) {
67                e.printStackTrace();
68        }
69
70 }
```

- Encrypt the message

With the public key in the directory and some configuration lines, the following code creates an encoded string based in the previously downloaded public key. This message can only be decrypted with the private key of the server.

```
1 File pubKeyFile = new File(rutaPkey);
2 DataInputStream dis;
3 byte[] encrypted;
4
5 dis = new DataInputStream(new FileInputStream(pubKeyFile));
6 byte[] keyBytes = new byte[(int) pubKeyFile.length()];
7 dis.readFully(keyBytes);
8 dis.close();
9 X509EncodedKeySpec keySpec = new X509EncodedKeySpec(keyBytes);
10 KeyFactory keyFactory;
11 keyFactory = KeyFactory.getInstance("RSA");
12 Cipher cipher = Cipher.getInstance("RSA/ECB/PKCS1PADDING");
13 RSAPublicKey publicKey = (RSAPublicKey)
keyFactory.generatePublic(keySpec);
14 cipher.init(Cipher.ENCRYPT_MODE, publicKey);
15
16 String usu_hash_enc =
bytesToHex(cipher.doFinal(hashUsuario.toString().getBytes("UTF-
8")));
```

- Send the encrypted message.

The encoded string is sent to the server with a HTTP request.

```
1 String url_insert = "http://mobilitapp.noip.me/insert.php";
2 JSONParser jsonParser = new JSONParser();
3 int success = 0;
4 List<NameValuePair> params = new ArrayList<NameValuePair>();
5
6 params.add(new BasicNameValuePair("usu_hash_enc",
usu_hash_enc));
7 JSONObject json = jsonParser.makeHttpRequest(url_create_regid,
"POST", params);
8
9 success = json.getInt(TAG_SUCCESS);
10
11 if (success == 1) {
12     result = "OK";
13     return result;
14 } else {
15     result = "KO";
16     return result;
17 }
```

The Class *JSONParser* will create the appropriate format for the HTTP call, and will also manage the response that the server gives to this request.

```java
1  public class JSONParser {
2      static InputStream is = null;
3      static JSONObject jObj = null;
4      static String json = "";
5      // constructor
6      public JSONParser() {     }
7      // function get json from url
8      // by making HTTP POST or GET mehtod
9      public JSONObject makeHttpRequest(String url, String method,
10             List<NameValuePair> params) {
11         // Making HTTP request
12         try {
13             // check for request method
14             if(method == "POST"){
15                 // request method is POST
16                 // defaultHttpClient
17                 DefaultHttpClient httpClient = new DefaultHttpClient();
18                 HttpPost = new HttpPost(url);
19                 httpPost.setEntity(new UrlEncodedFormEntity(params));
20                 Log.v("json:","url:"+httpPost.toString());
21
22                 HttpResponse httpResponse = httpClient.execute(httpPost);
23                 HttpEntity httpEntity = httpResponse.getEntity();
24                 is = httpEntity.getContent();
25             }else if(method == "GET"){
26                 // request method is GET
27                 DefaultHttpClient httpClient = new DefaultHttpClient();
28                 String paramString = URLEncodedUtils.format(params, "utf-8");
29                 url += "?" + paramString;
30                 Log.v("json:","url:"+url);
31                 HttpGet httpGet = new HttpGet(url);
32                 HttpResponse httpResponse = httpClient.execute(httpGet);
33                 HttpEntity httpEntity = httpResponse.getEntity();
34                 is = httpEntity.getContent();
35             }
36
37         } catch (UnsupportedEncodingException e) {
38             e.printStackTrace();
39         } catch (ClientProtocolException e) {
40             e.printStackTrace();
41         } catch (IOException e) {
42             e.printStackTrace();
43         }
44         try {
45             BufferedReader reader = new BufferedReader(new InputStreamReader(
46                     is, "iso-8859-1"), 8);
47             StringBuilder sb = new StringBuilder();
48             String line = null;
49             while ((line = reader.readLine()) != null) {
50                 sb.append(line + "\n");
51                 Log.v("json: ",line);
52             }
53             is.close();
54          //   json = sb.toString();
55             json = sb.toString().substring(0, sb.toString().length()-1);
56         } catch (Exception e) {
57             Log.e("Buffer Error", "Error converting result " + e.toString());
58         }
59         // try parse the string to a JSON object
60         try {
61             jObj = new JSONObject(json);
62         } catch (JSONException e) {
63             Log.e("JSON Parser", "Error parsing data " + e.toString());
64         }
65         // return JSON String
66         return jObj;
67     }
68 }
```

## 4.3. The decryption process

When the server receives this kind of HTTP requests, verifies they have the appropriate format, then uses the private key to decrypt the messages before saving the information to the database.

In the server, the PHP described next will first test if all the necessary fields are filled, then decrypt the message and makes the MySQL procedure needed on the database.

```php
1   <?php
2   // array for JSON response
3   $response = array();
4   //clave privada para decrypt
5   $fp = fopen("private.pem","r");
6   $privateKey = fread($fp,8192);
7   fclose ($fp);
8   $res = openssl_get_privatekey($privateKey);
9   // check for required fields
10  if (isset($_POST['usu_hash_enc']) && isset($_POST['reg_id_enc'])){
11
12      $usuhashenc = $_POST['usu_hash_enc'];
13      $regidenc = $_POST['reg_id_enc'];
14
15  function decryptar ($campo,$res){
16  openssl_private_decrypt(hex2bin($campo),$decrypted,$res);
17  return $decrypted;
18  }
19
20  $usuhash = decryptar($usuhashenc,$res);
21  $regid = decryptar($regidenc,$res);
22
23      // include db connect class
24      require_once __DIR__ . '/db_connect.php';
25      // connecting to db
26      $db = new DB_CONNECT();
27      // Insertar en tabla segmento y
28      $result = mysql_query("insert into tbl_usuarios (usu_hash,usu_regid)
value ('$usuhash','$regid') on duplicate key update usu_regid='$regid'");
29
30      // check if row inserted or not
31      if ($result) {
32          // successfully inserted into database
33          $response["success"] = 1;
34          $response["message"] = "Inserted";
35          // echoing JSON response
36          echo json_encode($response);
37      } else {
38          // faile:d to insert row
39          $response["success"] = 0;
40          $response["message"] = "Oops! An error occurred.";
41          // echoing JSON response
42          echo json_encode($response);
43      }
44  } else {
45      // required field is missing
46      $response["success"] = 0;
47      $response["message"] = "Required field(s) is missing";
48      // echoing JSON response
49      echo json_encode($response);
50  }
51  ?>
```

## 4.4. Performance Analysis

As stated earlier, this project uses a Raspberry Pi as web server and as a database manager that is why it is mandatory to perform a performance analysis due to limited hardware resources.

The highest limitation on the hardware is, theoretically, its CPU which runs at 700 MHz at most. This restriction on the processing power affects directly to the security implemented in the project as the encrypt/decrypt process requires a high use of processing resources. There are other limitations on the equipment as the RAM memory and the writing speed on the drive, these aspects should be properly studied and upgraded when the number of users increment. In the current proposal, these components will not create a bottleneck.

To find the limit of the CPU, we send an encrypted string every 10 seconds for about 40 minutes. This procedure will simulate a real scenario where all the users will upload their encrypted information around 1 defined hour.

Each character is decrypted and stored in the database. Then we change the length of the key used to encrypt and decrypt.

In Figure 10 and Figure 11 the results can be appreciated. It can be noticed a substantial difference between encrypting the information with 4096 bits or 2048 bits.

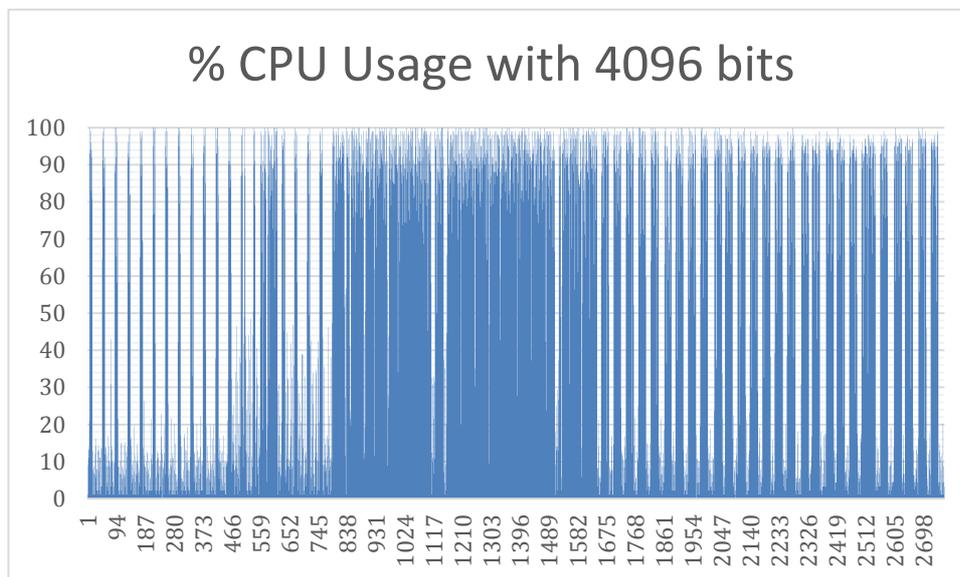

Figure 4-1: CPU graph with a key of 4096 bits

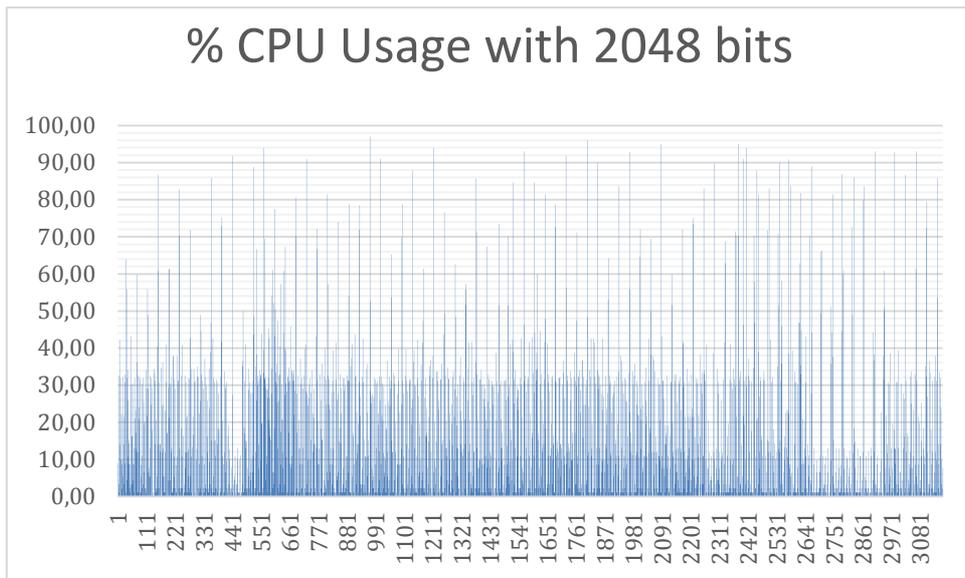

With the obtained results, the key length established for the security method applied is 2048 bits.

# Chapter 5
# Traffic Information

# 5. Traffic Information

One of the main features added to the MobilitApp is the real-time traffic information. The two main elements are traffic state and traffic incidences.
In this section, the tools used to provide this information are described.

## 5.1. Gathering the information

The real-time traffic state information is extracted from the Open data service administered by the Barcelona's City Hall [19]. This service presents the traffic state updated in a 5 minutes interval in a public access link which we call "dadestram.dat", the information showed is classified with the following format:

```
1#20150619130553#4#3
2#20150619130553#2#2
3#20150619130553#2#2
4#20150619130553#2#3
5#20150619130553#2#2
6#20150619130553#3#3
7#20150619130553#2#2
8#20150619130553#2#2
9#20150619130553#2#2
10#20150619130553#2#2#2
11#20150619130553#3#3
12#20150619130553#1#1
13#20150619130553#2#2
14#20150619130553#3#3
15#20150619130553#2#2
16#20150619130553#2#2
```

1#20080516103551#2#2. The first value, 1#, is the road section identifier. The following part is the date plus the hour, minutes and seconds. That is followed by a #2, which stands for current status (0 = no data, 1 = very fluid, 2 = fluid, 3 = dense, 4 = very dense, 5 = congested, 6 = closed). Finally, another value is shown which indicates the expected status in 15 minutes.

**Figure 5-1: "dadestrams.dat" format**

The other component necessary to show in the map the traffic information is the geo-coordinates of each of the roads described in "dadestrams.dat".
The Barcelona Open Data shows this information in an excel file called "trams.xls", with the next scheme:

**Figure 5-2: "trams.xls" format**

The first column contains the road section identifier, which will be the identification to find in "dadestram.dat", the second a text description of the road, while the third has a list of pairs latitude/longitude for each road.

With the information of this two files, one can build a map putting colours on the lines depending on the traffic state.

The algorithm designed is shown in Figure 5-3: Summary of how to present the traffic information.

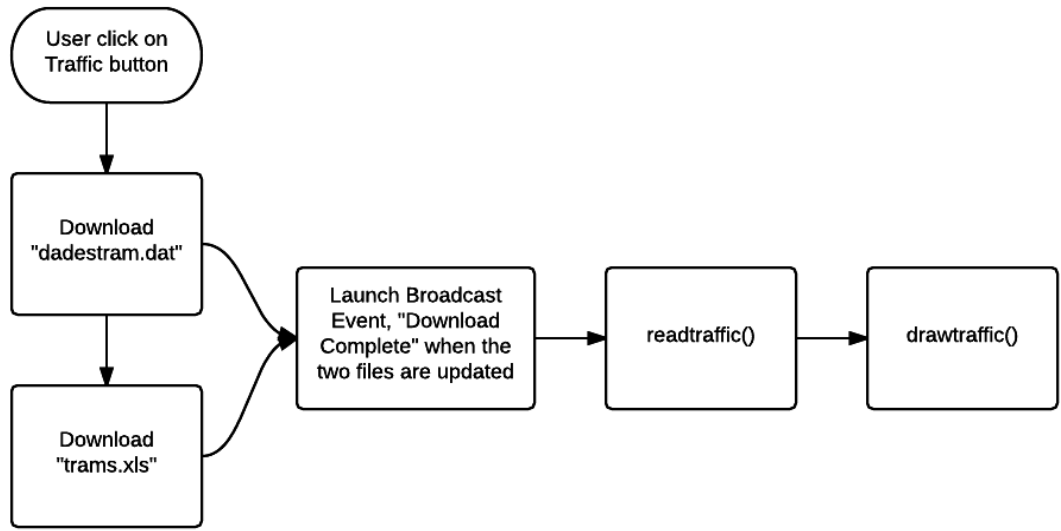

The functions readtraffic follows the code below:

```java
1 public readXlsCsv readTraffic(){
2 Log.v("traffic","Leer tráfico");
3 String rutaTrafico = Environment.DIRECTORY_DOWNLOADS +
"/MobilitApp/traffic";
4 String urlInfoTrams =
"http://www.bcn.cat/transit/dades/dadestrams.dat";
5 String urlTrams = "http://opendata.bcn.cat/opendata/en/descarrega-
fitxer?url=http%3a%2f%2fbismartopendata.blob.core.windows.net%2fopendata%
2fopendata%2fTRANSIT_RELACIO_TRAMS.xls&name=TRANSIT_RELACIO_TRAMS.xls";
6
7 try{
8                     //leer xls
9                     Double[][] coord;
10                    ExcelReader.RowConverter<Tramo> converter = (HSSFRow)
-> new Tramo(HSSFRow[0], HSSFRow[1],HSSFRow[2]);
11                    Log.v("xsl","crear workbook");
12                    ExcelReader<Tramo> reader =
ExcelReader.builder(Tramo.class)
13                          .converter(converter)
14                          .withHeader()
15                          .csvDelimiter(';')
16                          .sheets(1)
17                          .build();
18                    Log.v("xsl","creado worrkboook");
19
20                    List<Tramo> listaTramos;
21                    listaTramos =
reader.read(Environment.getExternalStoragePublicDirectory(Environment.DIR
ECTORY_DOWNLOADS).getAbsolutePath() + "/MobilitApp/traffic/trams.xls");
22
23
24                    //  crear objeto infotramos csv
25
26                    ExcelReader.RowConverter<infoTramo> converter2 =
(Row) -> new infoTramo(Row[0],Row[1],Row[2],Row[3]);
27                    Log.v("xsl","crear csv");
28                    ExcelReader<infoTramo> reader2 =
ExcelReader.builder(infoTramo.class)
29                          .converter(converter2)
30                          .csvDelimiter('#')
31                          .build();
32                    Log.v("xsl","creado csv");
33                    List<infoTramo> infoTramos;
34                    infoTramos =
reader2.read(Environment.getExternalStoragePublicDirectory(Environment.DI
RECTORY_DOWNLOADS).getAbsolutePath() +
"/MobilitApp/traffic/infotrams.csv");
35
36                    Log.v("xsl","infotramossize:"+infoTramos.size()+"");
37                    for(Tramo tramos : listaTramos) {
38                        // Log.v("xsl: ", tramos.coordenadas+"");
39
40                        if (tramos.coordenadas != null) {
41                            String[] pares =
tramos.coordenadas.split(",0");
42                            int i = 0;
43                            coord = new Double[pares.length][2];
44
45                            while (i < pares.length) {
46                                //quitar espacios en blanco
47                                //extraer latitude longitude
48                                pares[i] =
pares[i].replaceAll("\\s", "");
```

```
49                                        String[] latLong =
pares[i].split(",");
50
51                                        coord[i][0] =
Double.parseDouble(latLong[0]);
52                                        coord[i][1] =
Double.parseDouble(latLong[1]);
53                                        i++;
54                                    }
55                                    //guardar en objeto tramo
56                                    tramos.setCoord(coord);
57
58                        }
59                    }
60                    return new readXlsCsv(listaTramos, infoTramos);
61    }catch (Exception e) {
62        Log.e ("xsl",e.toString());
63        return null;
64    }
65 }
```

The map shows the lines with the corresponding colour with the next code:

```
1 public void onReceive(Context ctxt, Intent intent) {
2            readXlsCsv resultado = readTraffic();
3            List<Tramo> Tramos = resultado.getXLS();
4            List<infoTramo> infoTramos = resultado.getCSV();
5
6            LatLng latlng = null;
7            ArrayList<LatLng> latlngList;
8
9            //   int i =0;
10
11           for(Tramo tramos : Tramos) {
12               int i=0;
13               latlngList = new ArrayList<LatLng>();
14               if (tramos.getcoord() != null){
15
16                   while (i < tramos.getcoord().length) {
17                       // Log.v("xsl", "llenando lista vacia");
18                       latlng = new LatLng(tramos.getcoord()[i][1],
tramos.getcoord()[i][0]);
19                       latlngList.add(latlng);
20                       i++;
21                   }
22                   //  Log.v("xsl", "lista terminada de llenar");
23                   if (!latlngList.isEmpty()) {
24
25                       googleMap.addPolyline(new
PolylineOptions().addAll(latlngList).width(4).color(getTrafficColor(tramo
s,infoTramos)));
26                       iV2.setImageResource(R.drawable.leyendatrafico);
27                       ll2.setVisibility(View.VISIBLE);
28                       ll.setVisibility(View.GONE);
29                       hlv.setVisibility(View.GONE);
30
31                   } else {
32                       //  Log.v("xsl", "lista vacia");
33               }
34           }
35       }
```

### 5.1.1. Google Traffic

Google, and its Maps API are equipped with a layer which shows visual information of the traffic state. When the user asks for traffic information, this layer is also activated.

## 5.2. Incidences

The traffic incidences are provided by the Spanish Traffic Authority (DGT) [20]. The information is gathered in a public access web page with a 5 minutes interval.

The MobilitApp consults the webpage with the incidence information. The data is stored in a JSON format, then parsed and showed in the user's map. When the user click on the marker's box, the details of the incidence will appear in the screen as shown in figure 13.

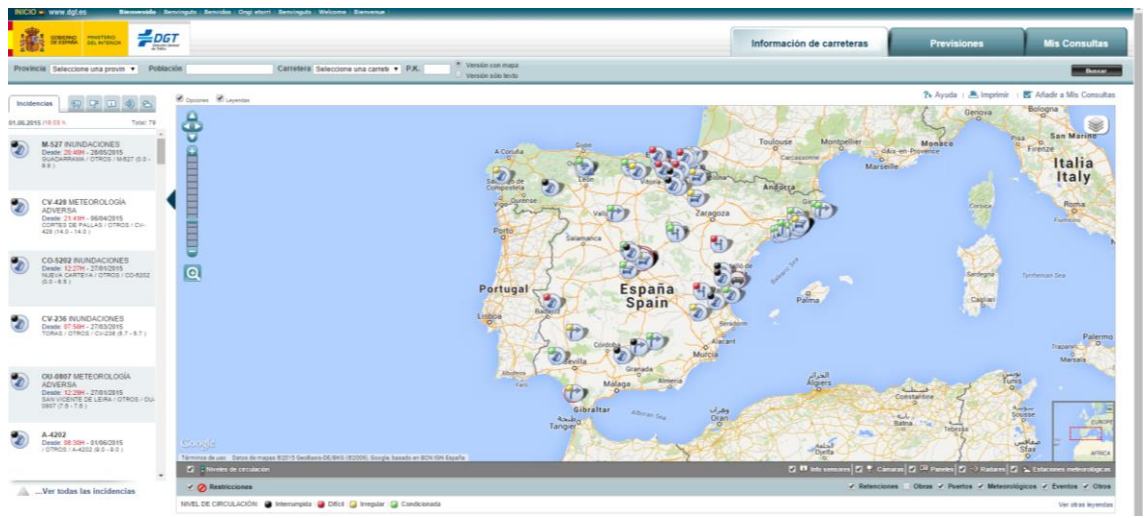

**Figure 5-4: DGT Web Page**

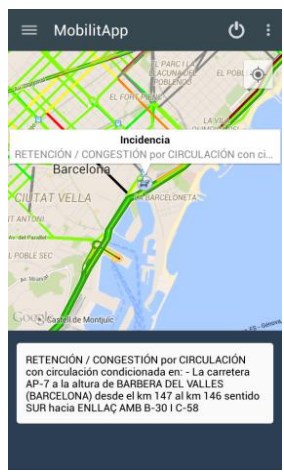

**Figure 5-5: The MobilitApp with the Traffic information on the map**

The following box shows the code designed to get and parse the information obtained from the DGT web page. The code creates the objects necessary to show the proper location, information and associated graphic on the user's screen.

```
1  JSONArray incidencias=new JSONArray();
2
3  try {
4      String urlTraficoIcons =
"http://infocar.dgt.es/etraffic/img/iconosIncitar/";
5      incidencias = new getIncidencias().execute().get();
6
7      for (int i = 0; i < incidencias.length(); i++) {
8          JSONObject c = incidencias.getJSONObject(i);
9          Log.v("json: ", "icono: " +urlTraficoIcons+
c.getString("icono"));
10          LatLng incidenciaCoord = new
LatLng(Double.parseDouble(c.getString("lat")),
Double.parseDouble(c.getString("lng")));
11          String textnoHtml =
Html.fromHtml(Html.fromHtml(c.getString("descripcion")).toString()).toStr
ing();
12          try {
13              Bitmap bmImg = Ion.with(getActivity())
14
.load(urlTraficoIcons+c.getString("icono")).asBitmap().get();
15              googleMap.addMarker(new MarkerOptions()
16                      .position(incidenciaCoord)
17                      .title("Incidencia")
18                      .snippet(textnoHtml)
19
.icon(BitmapDescriptorFactory.fromBitmap(bmImg))
20              );
21          } catch (InterruptedException e) {
22              e.printStackTrace();
23          } catch (ExecutionException e) {
24              e.printStackTrace();
25          }
26      }
28}
```

In line 5 the class *getIncidencias* is called, this class will make HTTP request to the DGT server to the get the incidences information. Android enforces to make this kind of operations in a class with the properties of a called "AsyncTask" to avoid interferences in the normal operation of any application, its definition is described below.

```java
1 public class getIncidencias extends AsyncTask<Void, Void, JSONArray> {
2 public getIncidencias() {
3
4 }
5
6 protected JSONArray doInBackground(Void... parameters) {
7
8       JSONArray incidencias = new JSONArray();
9       try {
10              //dibujar incidencias
11
12              String urlTraficoDgt =
"http://infocar.dgt.es/etraffic/BuscarElementos";
13
14              //guardar parametros de segment en un array
15              List<NameValuePair> params = new ArrayList<NameValuePair>();
16              params.add(new BasicNameValuePair("latNS",
"44.33956524809713"));
17              params.add(new BasicNameValuePair("longNS",
"30.1904296875"));
18              params.add(new BasicNameValuePair("latSW",
"26.745610382199022"));
19              params.add(new BasicNameValuePair("longSW", "-
39.287109375"));
20              params.add(new BasicNameValuePair("zoom", "5"));
21              params.add(new BasicNameValuePair("accion",
"getElementos"));
22              params.add(new BasicNameValuePair("Camaras", "false"));
23              params.add(new BasicNameValuePair("SensoresTrafico",
"false"));
24              params.add(new BasicNameValuePair("SensoresMeteorologico",
"false"));
25              params.add(new BasicNameValuePair("Paneles", "false"));
26              params.add(new BasicNameValuePair("IncidenciasRETENCION",
"true"));
27              params.add(new BasicNameValuePair("IncidenciasOBRAS",
"false"));
28              params.add(new
BasicNameValuePair("IncidenciasMETEOROLOGICA", "true"));
29              params.add(new BasicNameValuePair("IncidenciasPUERTOS",
"true"));
30              params.add(new BasicNameValuePair("IncidenciasOTROS",
"true"));
31              params.add(new BasicNameValuePair("IncidenciasEVENTOS",
"true"));
32              params.add(new BasicNameValuePair("niveles", "false"));
33              params.add(new BasicNameValuePair("caracter",
"acontecimiento"));
34
35              // Creating service handler class instance
36              ServiceHandler sh = new ServiceHandler();
37              // Making a request to url and getting response
38              String jsonStr = sh.makeServiceCall(urlTraficoDgt,
ServiceHandler.GET,params);
39              incidencias = new JSONArray(jsonStr);
40
41      } catch (Exception e) {

43
44      }
45      return incidencias;
46 }
47
48 }
```

Finally, in line ₃₆ another class is invoked: *ServiceHandler*. This class will manage the HTTP requests and replies. Constructing the HTTP request with all the parameters needed and placing the information replied by the DGT in a string variable.

```java
1 public class ServiceHandler {
2
3     String response = null;
4     public final static int GET = 1;
5     public final static int POST = 2;
6
7     public ServiceHandler() {
8
9     }
16     public String makeServiceCall(String url, int method) {
17         return this.makeServiceCall(url, method, null);
18     }
19
26     public String makeServiceCall(String url, int method,
27                                   List<NameValuePair>
params) {
28         try {
29             // http client
30             DefaultHttpClient httpClient = new
DefaultHttpClient();
31             HttpEntity httpEntity = null;
32             HttpResponse httpResponse = null;
34             // Checking http request method type
35             if (method == POST) {
36                 HttpPost httpPost = new HttpPost(url);
37                 // adding post params
38                 if (params != null) {
39                     httpPost.setEntity(new
UrlEncodedFormEntity(params));
40                 }
42                 httpResponse = httpClient.execute(httpPost);
44             } else if (method == GET) {
45                 // appending params to url
46                 if (params != null) {
47                     String paramString = URLEncodedUtils
48                             .format(params, "utf-8");
49                     url += "?" + paramString;
50                 }
51                 HttpGet httpGet = new HttpGet(url);
52
53                 httpResponse = httpClient.execute(httpGet);
55             }
56             httpEntity = httpResponse.getEntity();
57             response = EntityUtils.toString(httpEntity);
59         } catch (UnsupportedEncodingException e) {
60             e.printStackTrace();
61         } catch (ClientProtocolException e) {
62             e.printStackTrace();
63         } catch (IOException e) {
64             e.printStackTrace();
65         }
67         return response;
69     }
70 }
```

**Chapter 6**
**Extended Features in the MobilitApp**

# 6. Extended features

This chapter will describe at first, more new features added to the MobilitApp aiming to improve the user experience. Then will describe the methods used to spread the use of the application.

## 6.1. Presenting Results

A part of the focus of this project was to build mechanism to exploit the information that is being saved in the database.
One of the tools developed is a web form, where the transport authority can consult the database and filter its content applying the filters provided.

### 6.1.1. Consulting database

The queries to the database will be triggered through a web page (code in the annexes), that will call a PHP instance which contains the MySQL instructions necessaries to connect to the database, will then launch the queries and show the results in the requested format.

In the form, the user can filter the information by age, activity, date and time. Finally the results will be presented in a map (Figure 16) or in CSV format.

Age Between:
14

And:
40

Activity:
vehicle

Start:
12 June 2015 - 02:05 am

End:
14 June 2015 - 07:13 pm

Map          CSV

Figure 6-1: Form used to filter the mobility information.

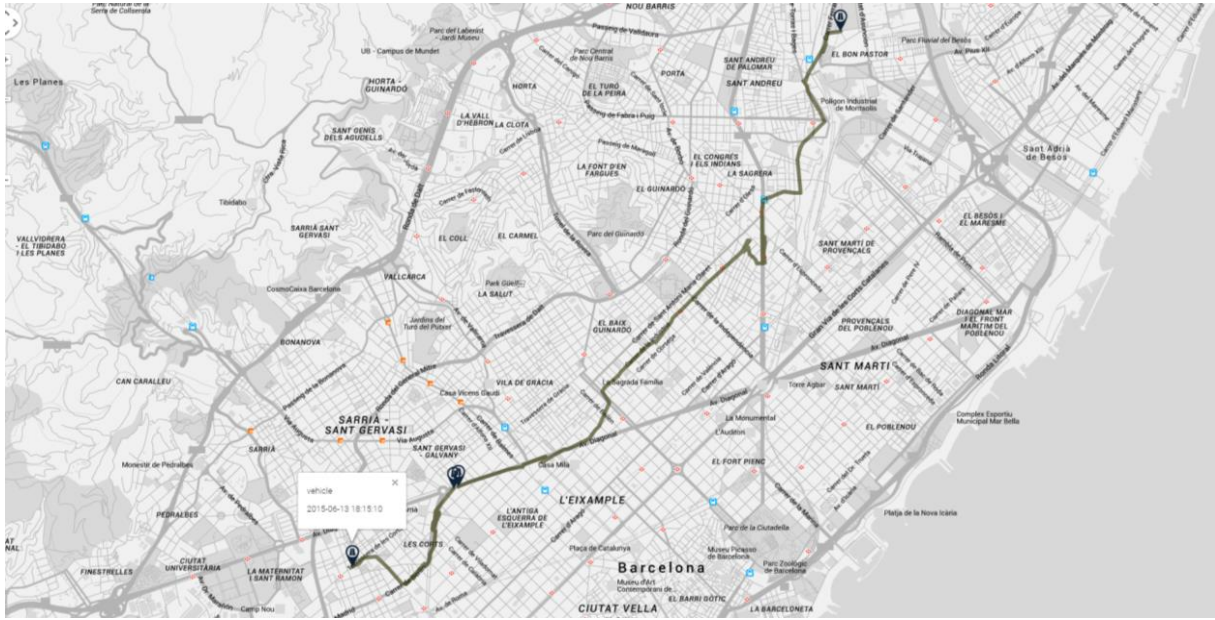

**Figure 6-2: Map showing the results.**

## 6.2. Silence Mode

The main two reasons to add this features are:

- When driving, the driver should not be distracted in any moment.
- When going on a public transportation vehicle, the noise produced by a mobile can bother the other passengers.

This mode is activated by default, but it is included in a menu option where it can be easily activated/deactivated.

When turned on, if the application detects that the user is moving in a vehicle, the terminal will be set to silence.

```
1 private void SilentMode() {
2
mAudioManager.setRingerMode(AudioManager.RINGER_MODE_SILENT);
3    }
```

This function is called when the application call the method "setActivity (String)" of the class "Segment". We use as a temporal value, a field in the SharedPreferences.

The algorithm follows:

- If the activity is vehicle, get the actual state and save it in the variable "estadoAnterior", then put the mobile in silence.
- When the detected activity is not vehicle put the mobile in the state previously saved in the shared preference variable.
- When detecting other activity don't make any change in the state of the terminal.

```
1 if ("vehicle".equals(activity) &&
prefes.getString("Silence","OFF").equalsIgnoreCase("ON")){
2       if (prefes.getInt("estadoAnterior",5)==5) {
3
4               switch (mAudioManager.getRingerMode()) {
5
6                       case 0:
7                               editor.putInt("estadoAnterior", 0);
8                               editor.commit();
9                               break;
10                      case 1:
11                              editor.putInt("estadoAnterior", 1);
12                              editor.commit();
13                              break;
14                      case 2:
15                              editor.putInt("estadoAnterior", 2);
16                              editor.commit();
17                              break;
18                      default:
19                              editor.putInt("estadoAnterior",
mAudioManager.getRingerMode());
20                              editor.commit();
21                              break;
22              }
23
24      }
25
26      SilentMode();
27
28}else{
29
30      if (prefes.getInt("estadoAnterior",5)!=5) {
31
32              mAudioManager.setRingerMode(prefes.getInt("estadoAnterior",
mAudioManager.getRingerMode()));
33              editor.putInt("estadoAnterior", 5);
34              editor.commit();
35
36      }else {
37
38              editor.putInt("estadoAnterior", 5);
39              editor.commit();
40
41      }
42      }
```

## 6.3. Help

This new component provide the user a brief guide on the app, describing the functionality of the main buttons on the screen.

The library ShowcaseView implements the graphical interfaces and animations of the added help. Figure 17 is a screenshot of the mobile when the users click on the "Help" button.

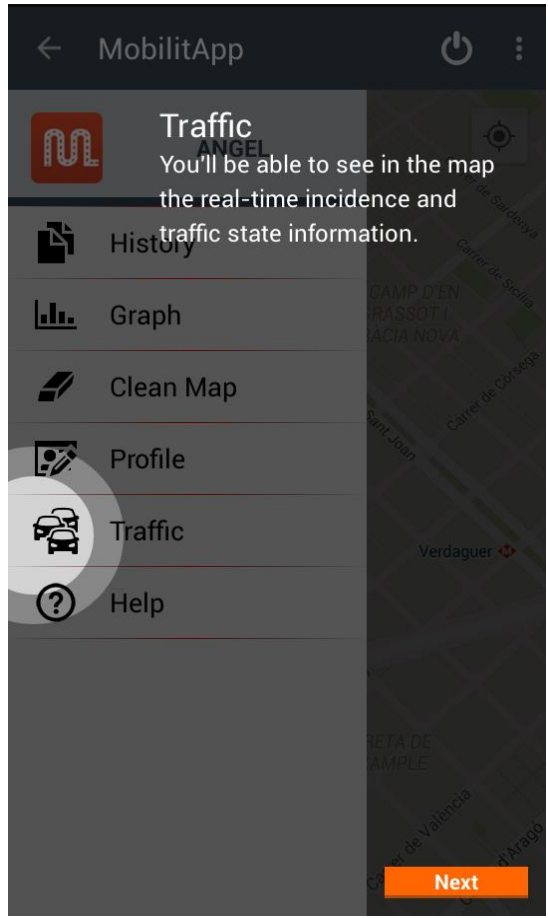

Figure 6-6-3: Help screen

This is the code necessary to build the help contents:

```java
1  public class Help extends Activity implements View.OnClickListener{
3      private ShowcaseView showcaseView;
4      private int contador=0;
5      private Target t1,t2,t3,t4,t5,t6,t7;
6
7      @Override
8      protected void onCreate(Bundle savedInstanceState) {
9          super.onCreate(savedInstanceState);
10         setContentView(R.layout.activity_help);
11
12         t1 = new ViewTarget(R.id.button,this);
13         t2 = new ViewTarget(R.id.button2,this);
14         t3 = new ViewTarget(R.id.button3,this);
15         t4 = new ViewTarget(R.id.button4,this);
16         t5 = new ViewTarget(R.id.button5,this);
17         t6 = new ViewTarget(R.id.button6,this);
18         t7 = new ViewTarget(R.id.button7,this);
22         showcaseView = new ShowcaseView.Builder(this,true)
23                     .setTarget(Target.NONE)
24                     .setOnClickListener(this)
25                     .setContentTitle(R.string.help)
26                     .setContentText(R.string.content_help)
27                     .setStyle(R.style.Transparencia)
28
29                     .build();
30         showcaseView.setButtonText
(getResources().getString(R.string.next));
34     }
37     @Override
38     public void onClick(View v) {
39
40         Log.v("aki", "contador" + contador);
41         switch (contador){
42             case 0:
43                 showcaseView.setShowcase(t1,true);
44
showcaseView.setContentTitle(getResources().getString(R.string.prefes));
45
showcaseView.setContentText(getResources().getString(R.string.text_prefes));
46                 break;
47             case 1:
48                 showcaseView.setShowcase(t2,true);
49
showcaseView.setContentTitle(getResources().getString(R.string.turn_off));
50
showcaseView.setContentText(getResources().getString(R.string.text_turn_off)
);
51                 break;
52             case 2:
53                 showcaseView.setShowcase(t3,true);
54
showcaseView.setContentTitle(getResources().getString(R.string.history_menu)
);
55
showcaseView.setContentText(getResources().getString(R.string.text_history_m
enu));
56                 break;
57             case 3:
58                 showcaseView.setShowcase(t4,true);
59
showcaseView.setContentTitle(getResources().getString(R.string.graph
_menu));
60
showcaseView.setContentText(getResources().getString(R.string.text_g
raph_menu));
61                 break;
```

```
62              case 4:
63                  showcaseView.setShowcase(t5,true);
64
showcaseView.setContentTitle(getResources().getString(R.string.clean
_menu));
65
showcaseView.setContentText(getResources().getString(R.string.text_c
lean_menu));
66                  break;
67              case 5:
68                  showcaseView.setShowcase(t6,true);
69
showcaseView.setContentTitle(getResources().getString(R.string.profi
le_menu));
70
showcaseView.setContentText(getResources().getString(R.string.text_p
rofile_menu));
71                  break;
72              case 6:  showcaseView.setShowcase(t7,true);
73                  showcaseView.setShowcase(t7,true);
74
showcaseView.setContentTitle(getResources().getString(R.string.traff
ic_menu));
75
showcaseView.setContentText(getResources().getString(R.string.text_t
raffic_menu));
76                  break;
77              case 7:
78                  this.finish();
79                  break;
80          }
81          contador++;
82      }
83 }
```

For using this code, there must be described the messages to be displayed in all languages supported by the application (Catalan, Spanish and English) inside "strings.xml".

## 6.4. Web page

To publicize the application, a static web page explaining the purpose of the application was designed: http://mobilitapp.noip.me/.The code is published in the annexes.

The web page is divided in the following sections:

- Header and Title
- Brief description of the application.
- Description of some of the features.
- Help screenshots to guide the users on the most common operations.
- Information of the team developing and supervising the application.

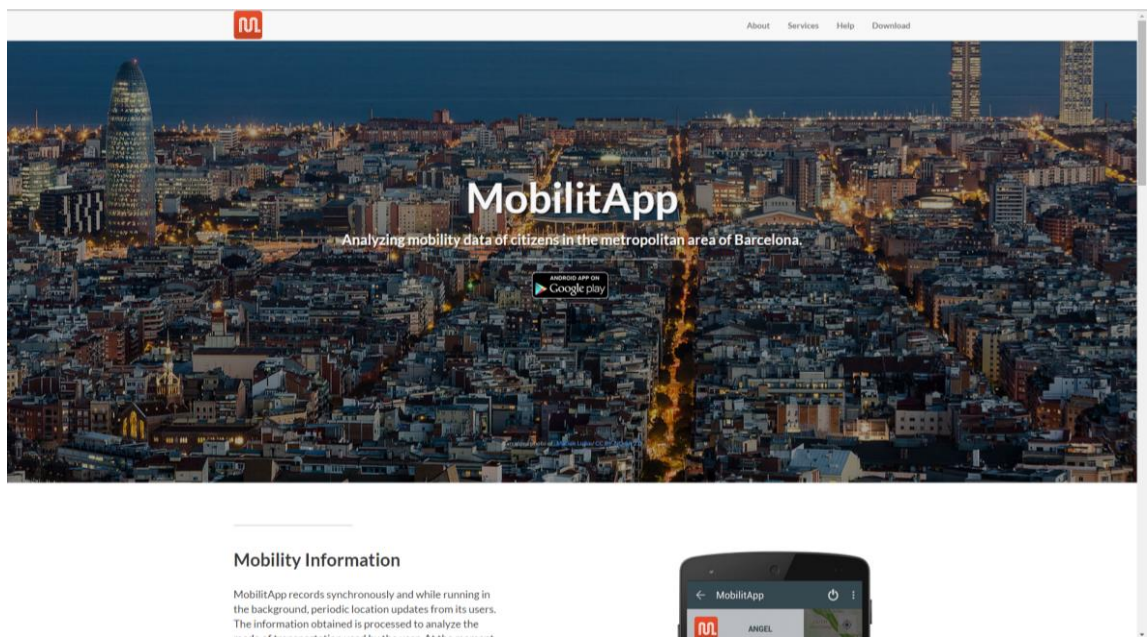

Figure 6-6-4: MobilitApp Webpage

## 6.5. Google Cloud Messaging

To establish a channel to connect the users with the server, this projects also add to the MobilitApp support to Google Cloud Messaging service.

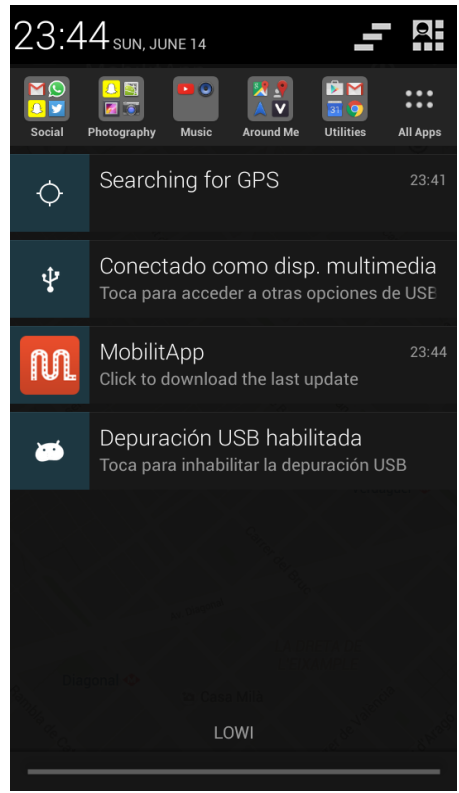

With the current code, we send a message to any user in the database with a personalized title and message. When the user clicks on the message, the message will redirect the mobile to the webpage of the MobilitApp. We use this service to warn the user when a new version of the application were available.

The code for handling this events involves 3 additional classes, its code is provided in the annexes:
- GcmBroadcastReceiver
- GCMClientManager
- GcmMessageHandler

Further, this code should be placed in the "onCreate ()" method of the MainActivity class, to register the user in the database.

```
1 private GCMClientManager pushClientManager;
2 String PROJECT_NUMBER = "443113651895",id="";
3
4 pushClientManager = new GCMClientManager(this, PROJECT_NUMBER);
5 pushClientManager.registerIfNeeded(new
GCMClientManager.RegistrationCompletedHandler() {
6     @Override
7     public void onSuccess(String registrationId, boolean
isNewRegistration) {
8             // SEND async device registration to your back-end server
9             // linking user with device registration id
10            // POST https://my-back-
end.com/devices/register?user_id=123&device_id=abc
11            try {
12                    String result = new SaveDatabase(ruta, registrationId,
prefs.getString("id", "no_id"), MainActivity.this).execute().get();
13                    if (result.equalsIgnoreCase("OK")) {
14                    //reg_id saved correctly.
15                    } else {
16                    //reg_id not saved in the database
17                    }
18            } catch (InterruptedException e) {
19                    e.printStackTrace();
20            } catch (ExecutionException e) {
21                    e.printStackTrace();
22            }
23     }
24
25     @Override
26     public void onFailure(String ex) {
27             super.onFailure(ex);
28             // If there is an error registering, don't just keep trying to
register.
29             // Require the user to click a button again, or perform
30             // exponential back-off when retrying.
31     }
32});
```

On 12 the code calls to the class "SaveDatabase", this call will trigger a constructor
of the class that sends an HTTP request to the server with only the registration ID
of the google service and one unique identifier of the user.

The encryption/decryption process is handled just as when saving a segment,
described in section IV.

Then, this code is added in the manifest of the application to register the classes
previously defined.

```
<receiver
        android:name="com.mobi.mobilitapp.GcmBroadcastReceiver"
        android:permission="com.google.android.c2dm.permission.SEND" >
        <intent-filter>
                <action
android:name="com.google.android.c2dm.intent.RECEIVE" />
                <category android:name="com.mobi.mobilitapp" />
        </intent-filter>
</receiver>
<service android:name="com.mobi.mobilitapp.GcmMessageHandler" />
```

## 6.6.  Google Play Store

Once the development was finished, we decide to publish the application to Google Play. This is Google's official store and portal for Android apps, games and other content for your Android-powered phone or tablet, where one can upload the APK of an application.

At first we had to create a developer account with a cost of $20 then we had to adapt the application structure to the store standards. These are some of the changes done in order to fulfil the requirements, and the procedure followed to upload the application:

- Change the name of the package:
  In the store, packages names as "com.example.*" are not allowed. The package name of the app became "com.mobi.mobilitapp".

- Create a keystore in the developer equipment.
  Upload applications signed with the debug keystore is not permitted. Then, we create a new keystore associated to a release key.

- Create a new project in Google API console.
  The change of the name on the package name implied the indicated adjustmen. This step was crucial as our activity recognition depends directly on the Google APIs.  The creation of a new project implied a new set of API keys that were modified based on the previous project.

After all the modifications were done, the APK was uploaded successfully to the store, Figure 6-6 and Figure 6-7 shows a screenshot of the store profile given for the application and some statistics respectively,

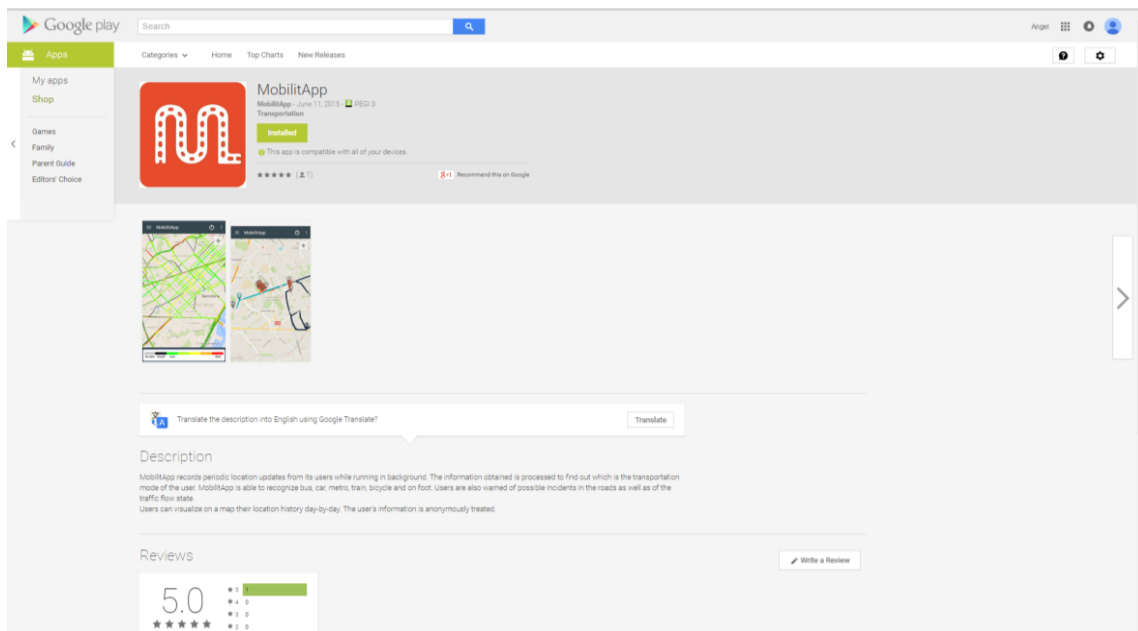

Figure 6-6-6: MobilitApp Play Store Page

| APP NAME | PRICE | CURRENT / TOTAL INSTALLS | AVG. RATING / TOTAL # | CRASHES & ANRS | LAST UPDATE | STATUS |
|---|---|---|---|---|---|---|
| MobilitApp 2.3 | Free | 12 / 12 | ★ 5.00 / 1 | — | Jun 11, 2015 | Published |

Page **1** of **1**

**Figure 6-7: MobilitApp Statistics of use.**

# Chapter 7
# Conclusions and future work

# 7. Conclusions and Future Work

With the information saved in a flexible format, as it is a relational database, the possibilities of new applications are enormous, some of them could be:

- To send warnings to the user in real-time when detecting an incidence around. The Google Cloud Messaging service established in section 6.5 can be used for this purposes.

- Predict and detect "usual routes" for each user, with it the application could propose alternative routes in case of any incidence.

- With the "usual route" defined, the city can be divided in geographic zones where one can send messages or actions to a group of users with similar tendencies. Then, having a significant amount of samples, a user can be modelled with a pattern, which can be used to warn about incidences and provide alternative routes to the "usual route" when an incidence is detected.

- The improvements on the activity detection algorithms. In the tests done the last few days with some of the users we have detected some punctual errors which generates abrupt changes on the mobility patterns.

- Join the work done in this project with [21] where the location is obtained with the raw values of some sensors in the terminal, mostly the accelerometer. And to use a technique where the positions are obtained based on the principle that a mobile uses more power when is far from a station and less power when get closer. Measuring the power of the terminal in each moment and knowing in advance the location of the closest GSM, Wi-Fi stations, one can triangle the location as stated in ---

We have already "open the gate" to this procedures, saving the power of the GSM power and Wi-Fi power sensors when sampling the locations.

It should be emphasized that the tools used in the development of this project focus on saving resources (money, energy). When the application reaches a wider audience, the tools must be adapted to the new needs which means higher processing power, a more reliable database system and improve the security methods.

# Chapter 8
# Annexes

# 8. Annexes

## 8.1. Server Installation Instructions

The server will run over an LAMP server supported by a Raspberry Pi hardware.
For the installation of the server, it is assumed that the Raspberry Pi main components (operating system: Raspbian, updates) are already installed [22], it is properly connected to internet and the SSH access is enabled [17].

Raspbian is a free operating system based on Debian optimized for the Raspberry Pi hardware, this is the software recommended for the server Linux distribution, and more configurations are described below.
Next, we will follow the other components necessary for the server: Apache, MySQL and PHP.

### 8.1.1. Linux configurations

We won't use the graphical interface of the Raspberry Pi that is why the following command is necessary to allocate less memory to the graphics. This command allocate only 32 MB to the graphics.

```
> sudo cp /boot/arm224_start.elf /boot/start.elf
```

The network should also be configured, the server needs a static IP configuration. This is an example of the content in the file /etc/network/interfaces having an static IP:

```
iface eth0 inet static

address 192.168.1.4

netmask 255.255.255.0

gateway 192.168.1.1
```

In the router, the traffic from/to port 80 and 22 should be redirected to this static IP. This configuration varies depending on the router.
Another aspect to configure is the DNS. In the project we use a public and free DNS server on Internet (NoIP), which is compatible with the router. Again this configuration depends on the router and the internet provider.

### 8.1.2. Installing more components

The following code will install the apache web service:

```
>sudo apt-get update
>sudo apt-get upgrade
>sudo apt-get install apache2
```

Then, the MySQL:

```
>sudo apt-get install MySQL-server
```

During the install there is a prompt request for a password. The password is for the MySQL root user.

Finally install PHP:

```
>sudo apt-get install php5
>sudo apt-get install php5-mysql
```

Once the setup is complete one can access the web page by pointing the browser to the router IP address or DNS entry. You should get a page back stating that it works, but that there is no content loaded.

To test that the webserver and PHP are working correctly then delete the file /var/www/index.html and create a file /var/www/index.php with a helloworld in php and check that it has the right content.

## 8.2. Google Cloud Messaging Classes

These 3 classes handle the Google Cloud Messaging service.

`GcmBroadcastReceiver`, creates a broadcast receiver which will be triggered when the GCM event is received in the terminal.

```java
public class GcmBroadcastReceiver extends
WakefulBroadcastReceiver {
    // Receives the broadcast directly from GCM service
    @Override
    public void onReceive(Context context, Intent intent) {
        // Explicitly specify that GcmMessageHandler will handle
the intent.
        ComponentName comp = new
ComponentName(context.getPackageName(),
                GcmMessageHandler.class.getName());
        // Start the service, keeping the device awake while it
is executing.
        startWakefulService(context,
(intent.setComponent(comp)));
        // Return successful
        setResultCode(Activity.RESULT_OK);
    }
}
```

GCMClientManager, class that creates the link with the API, producing the registration ID. Unique identifier of each terminal.

```java
public class GCMClientManager {
    // Constants
    public static final String TAG = "GCMClientManager";
    public static final String EXTRA_MESSAGE = "message";
    public static final String PROPERTY_REG_ID = "registration_id";
    private static final String PROPERTY_APP_VERSION = "appVersion";
    private final static int PLAY_SERVICES_RESOLUTION_REQUEST = 9000;
    // Member variables
    private GoogleCloudMessaging gcm;
    private String regid;
    private String projectNumber;
    private Activity activity;

    public static abstract class RegistrationCompletedHandler {
        public abstract void onSuccess(String registrationId, boolean
isNewRegistration);
        public void onFailure(String ex) {
            // If there is an error, don't just keep trying to
register.
            // Require the user to click a button again, or perform
            // exponential back-off.
            Log.e(TAG, ex);
        }
    }
```

```java
23
24     public GCMClientManager(Activity activity, String projectNumber) {
25         this.activity = activity;
26         this.projectNumber = projectNumber;
27         this.gcm = GoogleCloudMessaging.getInstance(activity);
28     }
29
30     // Register if needed or fetch from local store
31     public void registerIfNeeded(final RegistrationCompletedHandler
handler) {
32         if (checkPlayServices()) {
33
34
35             registerInBackground(handler);
36
37
38         } else { // no play services
39             Log.i(TAG, "No valid Google Play Services APK found.");
40         }
41     }
42
43     /**
44      * Registers the application with GCM servers asynchronously.
45      * <p>
46      * Stores the registration ID and app versionCode in the
application's
47      * shared preferences.
48      */
49     private void registerInBackground(final
RegistrationCompletedHandler handler) {
50         new AsyncTask<Void, Void, String>() {
51             @Override
52             protected String doInBackground(Void... params) {
53                 try {
54                     if (gcm == null) {
55                         gcm =
GoogleCloudMessaging.getInstance(getContext());
56                     }
57                     regid = gcm.register(projectNumber);
58                     Log.i(TAG, regid);
59
60                     // Persist the regID - no need to register again.
61                     storeRegistrationId(getContext(), regid);
62
63                 } catch (IOException ex) {
64                     // If there is an error, don't just keep trying to
register.
65                     // Require the user to click a button again, or
perform
66                     // exponential back-off.
67                     handler.onFailure("Error :" + ex.getMessage());
68                 }
69                 return regid;
70             }
71
72             @Override
73             protected void onPostExecute(String regId) {
74                 if (regId != null) {
75                     handler.onSuccess(regId, true);
76                 }
77             }
78         }.execute(null, null, null);
79     }
80
81
```

```java
89      private String getRegistrationId(Context context) {
90          final SharedPreferences prefs = getGCMPreferences(context);
91          String registrationId = prefs.getString(PROPERTY_REG_ID, "");
92          if (registrationId.isEmpty()) {
93              Log.i(TAG, "Registration not found.");
94              return "";
95          }
96
97          // Check if app was updated; if so, it must clear the
registration ID
98          // since the existing regID is not guaranteed to work with the
new
99          // app version.
100         int registeredVersion = prefs.getInt(PROPERTY_APP_VERSION,
Integer.MIN_VALUE);
101         int currentVersion = getAppVersion(context);
102         if (registeredVersion != currentVersion) {
103             Log.i(TAG, "App version changed.");
104             return "";
105         }
106         return registrationId;
107     }
108
109     /**
110      * Stores the registration ID and app versionCode in the
application's
111      * {@code SharedPreferences}.
112      *
113      * @param context application's context.
114      * @param regId registration ID
115      */
116     private void storeRegistrationId(Context context, String regId) {
117         final SharedPreferences prefs = getGCMPreferences(context);
118         int appVersion = getAppVersion(context);
119         Log.i(TAG, "Saving regId on app version " + appVersion);
120         SharedPreferences.Editor editor = prefs.edit();
121         editor.putString(PROPERTY_REG_ID, regId);
122         editor.putInt(PROPERTY_APP_VERSION, appVersion);
123         editor.commit();
124     }
125
126     /**
127      * @return Application's version code from the {@code
PackageManager}.
128      */
129     private static int getAppVersion(Context context) {
130         try {
131             PackageInfo packageInfo = context.getPackageManager()
132                     .getPackageInfo(context.getPackageName(), 0);
133             return packageInfo.versionCode;
134         } catch (NameNotFoundException e) {
135             // should never happen
136             throw new RuntimeException("Could not get package name: "
+ e);
137         }
138     }
139
140     private SharedPreferences getGCMPreferences(Context context) {
141         // This sample app persists the registration ID in shared
preferences, but
142         // how you store the regID in your app is up to you.
143         return
getContext().getSharedPreferences(context.getPackageName(),
144                 Context.MODE_PRIVATE);
145     }
146
```

```
147    /**
148     * Check the device to make sure it has the Google Play Services
APK. If
149     * it doesn't, display a dialog that allows users to download the
APK from
150     * the Google Play Store or enable it in the device's system
settings.
151     */
152    private boolean checkPlayServices() {
153        int resultCode =
GooglePlayServicesUtil.isGooglePlayServicesAvailable(getContext());
154        if (resultCode != ConnectionResult.SUCCESS) {
155            if
(GooglePlayServicesUtil.isUserRecoverableError(resultCode)) {
156                GooglePlayServicesUtil.getErrorDialog(resultCode,
getActivity(),
157                    PLAY_SERVICES_RESOLUTION_REQUEST).show();
158            } else {
159                Log.i(TAG, "This device is not supported.");
160            }
161            return false;
162        }
163        return true;
164    }
165
166    private Context getContext() {
167        return activity;
168    }
169
170    private Activity getActivity() {
171        return activity;
172    }
173}
```

Is in charge of creating the action to be performed when the application receives a GCM message.

```
1  public class GcmMessageHandler extends IntentService {
2      public static final int MESSAGE_NOTIFICATION_ID = 435345;
3
4      public GcmMessageHandler() {
5          super("GcmMessageHandler");
6      }
7
8      @Override
9      protected void onHandleIntent(Intent intent) {
10         // Retrieve data extras from push notification
11         Bundle extras = intent.getExtras();
12         GoogleCloudMessaging gcm =
GoogleCloudMessaging.getInstance(this);
13         // The getMessageType() intent parameter must be the intent you
received
14         // in your BroadcastReceiver.
15         String messageType = gcm.getMessageType(intent);
16         // Keys in the data are shown as extras
17         String title = extras.getString("title");
18         String body = extras.getString("body");
19         // Create notification or otherwise manage incoming push
20         createNotification(title, body);
21         // Log receiving message
22         Log.i("GCM", "Received : (" + messageType + ")  " +
extras.getString("title"));
```

```
23          // Notify receiver the intent is completed
24          GcmBroadcastReceiver.completeWakefulIntent(intent);
25      }
26
27      // Creates notification based on title and body received
28      private void createNotification(String title, String body) {
29          Context context = getBaseContext();
30          NotificationCompat.Builder mBuilder = new
NotificationCompat.Builder(context)
31
.setSmallIcon(R.drawable.ic_logo).setContentTitle(title)
32              .setAutoCancel(true)
33              .setContentText(body);
34
35          NotificationManager mNotificationManager =
(NotificationManager) context
36              .getSystemService(Context.NOTIFICATION_SERVICE);
37
38          Intent notificationIntent = new Intent(Intent.ACTION_VIEW);
39
notificationIntent.setData(Uri.parse("http://mobilitapp.noip.me"));
40          PendingIntent pending = PendingIntent.getActivity(this, 0,
notificationIntent, PendingIntent.FLAG_UPDATE_CURRENT);
41          mBuilder.setContentIntent(pending);
42
43          mNotificationManager.notify(MESSAGE_NOTIFICATION_ID,
mBuilder.build());
44
45
46
47      }
48
49 }
```

## 8.3. Web Page Code

This is the code of the form used to filter the database information: "consult.php"

```html
<!DOCTYPE html>
<html>
<head>
<meta content="text/html; charset=UTF-8" http-equiv="Content-Type">
<meta charset="utf-8">
  <meta name="viewport" content="width=device-width, initial-scale=1">
<link rel="shortcut icon" href="logoo.ico" type="image/x-icon" />

<script>
function showUser() {
    var str=document.getElementById('q').value;
    if (str == "") {
        document.getElementById("txtHint").innerHTML = "";
        return;
    } else {
        if (window.XMLHttpRequest) {
            // code for IE7+, Firefox, Chrome, Opera, Safari
            xmlhttp = new XMLHttpRequest();
        } else {
            // code for IE6, IE5
            xmlhttp = new ActiveXObject("Microsoft.XMLHTTP");
        }
        xmlhttp.onreadystatechange = function() {
            if (xmlhttp.readyState == 4 && xmlhttp.status == 200) {
                document.getElementById("txtHint").innerHTML = xmlhttp.responseText;
            }
        }
        xmlhttp.open("GET","getLocations.php?q="+str,true);
        xmlhttp.send();
    }
}

function validate(evt) {
  var theEvent = evt || window.event;
  var key = theEvent.keyCode || theEvent.which;
  key = String.fromCharCode( key );
  var regex = /[0-9]|\./;
  if( !regex.test(key) ) {
    theEvent.returnValue = false;
    if(theEvent.preventDefault) theEvent.preventDefault();
  }
}
  function validateForm() {
    var x = document.forms["myForm"]["edad"].value;
    var x2 = document.forms["myForm"]["edad2"].value;
    var x3 = document.forms["myForm"]["actividad"].value;

    var x5 = document.getElementById('dtp_input2').value;
    var x4 = document.getElementById('dtp_input1').value;
```

```
      if (x == null || x == "" || x2 == null || x2 == "" ||  x3 ==
null || x3 == "" || x3 == ":" || x4 == "" || x5 == "" ) {
          alert("All fields must be filled out");
          return false;
      }
}

</script>
<link href="css/bootstrap.min.css" rel="stylesheet">

    <link href="css/font-awesome.min.css" rel="stylesheet">
    <link href="css/main.css" rel="stylesheet">
    <link href="css/animate.css" rel="stylesheet">
    <link href="css/responsive.css" rel="stylesheet">

    <link href="css/bootstrap-datetimepicker.min.css"
rel="stylesheet" media="screen">

</head>
<body>

 <?php
 $con =
mysqli_connect('localhost','mobilitapp','Mobilitapp2015','mobilita
pp');
if (!$con) {
    die('Could not connect: ' . mysqli_error($con));
}

$sql="SELECT * FROM tbl_usuarios";
$result = mysqli_query($con,$sql);

?>

<div class="container">

<div class="jumbotron">

<h1>MobilitApp</h1>

</div>

<div class="row">

    </div>
```

```html
        <form class="form-horizontal" name="myForm" method="post"
onsubmit="return validateForm()" action='getLocations.php'>

        <div class="form-group">

        <!--    <select class="col-sm-2" name="edad" id='q'>

                <option value="">Edad entre:</option>

                <option value="14">14</option>
                <option value="15">15</option>
                <option value="30">30</option>

            </select>
            -->

            <p>Age Between:<br>
             <input class="col-sm-12 col-md-6 col-xs-12"
default=14 type='number' onkeypress='validate(event)' name="edad"
size="6" name="age" min="14" max="99" /> </p>
                <br><br>

            And: <br>
            <input class="col-sm-12 col-md-6 col-xs-12"
type='number' onkeypress='validate(event)' name="edad2" size="6"
name="age" min="14" max="99"/>

        </div>
Activity:

        <div class="form-group">

            <select class="col-sm-12 col-md-6 col-xs-12"
name="actividad" id='r'>

                <option value="">:</option>
                <option value="All">All</option>

<?php
$sql2="SELECT distinct seg_activity FROM tbl_Segmento";
$result = mysqli_query($con,$sql2);

    while ($row = mysqli_fetch_array($result)) {

                $id = $row['seg_activity'];
                //$name = $row['name'];
                echo "<option value=".$id.">".$id."</option>";
        }
?>
```

```html
</select>
        </div>

        <div class="form-group">

                Start:
                <div class="input-group date form_datetime col-md-
6 col-sm-12 col-xs-12" data-date="1979-09-16T05:25:07Z" data-date-
format="dd MM yyyy - HH:ii p" data-link-field="dtp_input1" >
                        <input class="form-control" size="16"
type="text" value="" readonly>
                        <span class="input-group-addon"><span
class="glyphicon glyphicon-remove"></span></span>
                        <span class="input-group-addon"><span
class="glyphicon glyphicon-th"></span></span>
                </div>
                <input type="hidden" id="dtp_input1" value=""
name="dtp_input1" /><br/>

        </div>

        <div class="form-group">

                End:
                <div class="input-group date form_datetime col-md-
6 col-sm-12 col-xs-12" data-date="1979-09-16T05:25:07Z" data-date-
format="dd MM yyyy - HH:ii p" data-link-field="dtp_input2">
                        <input class="form-control" size="16"
type="text" value="" readonly>
                        <span class="input-group-addon"><span
class="glyphicon glyphicon-remove"></span></span>
                        <span class="input-group-addon"><span
class="glyphicon glyphicon-th"></span></span>
                </div>
                <input type="hidden" id="dtp_input2" value=""
name="dtp_input2" /><br/>

        </div>

        <div class="form-group">

            <button type="submit" name="submit_map"
formaction="getLocations.php" value="submit 1" class="btn btn-
default col-md-3 col-xs-6">Map</button>
            <button type="submit " name="submit_csv"
formaction="getCSV.php" value="submit 2" class="btn btn-default
col-md-3 col-xs-6">CSV</button>
        </div>

    </form>
</div>
```

```html
<script type="text/javascript" src="js/jquery-1.9.1.min.js"
charset="UTF-8"></script>
<script type="text/javascript" src="js/bootstrap.min.js"></script>
<script type="text/javascript" src="js/bootstrap-
datetimepicker.js" charset="UTF-8"></script>
<script type="text/javascript" src="js/locales/bootstrap-
datetimepicker.fr.js" charset="UTF-8"></script>
<script type="text/javascript">
    $('.form_datetime').datetimepicker({
        //language:  'fr',
        weekStart: 1,
        todayBtn:  1,
        autoclose: 1,
        todayHighlight: 1,
        startView: 2,
        forceParse: 0,
        showMeridian: 1
    });
    $('.form_date').datetimepicker({
        language:  'fr',
        weekStart: 1,
        todayBtn:  1,
        autoclose: 1,
        todayHighlight: 1,
        startView: 2,
        minView: 2,
        forceParse: 0
    });
    $('.form_time').datetimepicker({
        language:  'fr',
        weekStart: 1,
        todayBtn:  1,
        autoclose: 1,
        todayHighlight: 1,
        startView: 1,
        minView: 0,
        maxView: 1,
        forceParse: 0
    });
</script>

</body>
</html>
```

As it can be noticed, the contain of the filters are sent to other page which is in charge of launch the queries and present the results on the map or on a CSV file.

"getLocations.php" to show the markers on the map

```html
<!DOCTYPE html>
<html>
<head>
<link href="css/main.css" rel="stylesheet">
<style>
table {
    width: 100%;
    border-collapse: collapse;
}

table, td, th {
    border: 1px solid black;
    padding: 5px;
}

th {text-align: left;}
</style>
</head>
 <script type="text/javascript">
function addMarker(lat, lng, act,date,time) {
 var image = '/images/'+act+'.png';

 map.addMarker({
        lat: lat,
        lng: lng,
        icon: image,
        verticalAlign: 'bottom',
        horizontalAlign: 'center',
        backgroundColor: '#3e8bff',
        infoWindow: {
   content: '<p>'+act+'</p>'+'<p>'+date+' '+time+'</p>'
}
    });

 }
  function drawPolylines(path,name) {

switch(name) {
    case "still":
        var color = '#FF00FF'
        break;
    case "on_foot":
        var color = '#FF9900'
        break;
    case "vehicle":
        var color = '#333300'
        break;
    case "bicycle":
        var color = '#00FF00'
        break;
    case "bus":
        var color = '#0099CC'
        break;
}
```

```javascript
  map.drawPolyline({
  path: path,
  strokeColor: color,
  strokeOpacity: 0.6,
  strokeWeight: 6
});

 }

 function initMap() {
    map = new GMaps({
        el: '#gmap',
        lat: 41.400971,
        lng: 2.165102,
        scrollwheel:true,
        zoom: 12,
        zoomControl : true,
        panControl : true,
        streetViewControl : false,
        mapTypeControl: false,
        overviewMapControl: false,
        clickable: false
    });

        var styles = [

    {
        "featureType": "road",
        "stylers": [
        { "color": "#b4b4b4" }
        ]
    },{
        "featureType": "water",
        "stylers": [
        { "color": "#d8d8d8" }
        ]
    },{
        "featureType": "landscape",
        "stylers": [
        { "color": "#f1f1f1" }
        ]
    },{
        "elementType": "labels.text.fill",
        "stylers": [
        { "color": "#000000" }
        ]
    },{
        "featureType": "poi",
        "stylers": [
        { "color": "#d9d9d9" }
        ]
```

```
        },{
            "elementType": "labels.text",
            "stylers": [
            { "saturation": 1 },
            { "weight": 0.1 },
            { "color": "#000000" }
            ]
        }
        ];

    map.addStyle({
        styledMapName:"Styled Map",
        styles: styles,
        mapTypeId: "map_style"
    });

    map.setStyle("map_style");

 <?php
if(isset($_POST['edad'])){ $edad = $_POST['edad']; }
if(isset($_POST['edad2'])){ $edad2 = $_POST['edad2']; }
if(isset($_POST['actividad'])){ $actividad = $_POST['actividad'];
}
if(isset($_POST['dtp_input1'])){ $fecha1 = $_POST['dtp_input1'];
}
if(isset($_POST['dtp_input2'])){ $fecha2 = $_POST['dtp_input2']; }

$date1 = date( 'Y-m-d', strtotime($fecha1));
$date2 = date( 'Y-m-d', strtotime($fecha2));

$time1 = date( 'YYYY-mm-dd', strtotime($fecha1));
$time2 = date( 'hh-ii-ss', strtotime($fecha2));

$con =
mysqli_connect('localhost','mobilitapp','Mobilitapp2015','mobilita
pp');
if (!$con) {
    die('Could not connect: ' . mysqli_error($con));
}

if ($actividad == "All"){
$sql="select
a.loc_latitude,a.loc_longitude,a.loc_time,a.loc_date,b.seg_activit
y, b.seg_id from tbl_Location a JOIN tbl_Segmento b ON a.seg_id =
b.seg_id JOIN tbl_usuarios d ON b.usu_hash = d.usu_hash where
YEAR(CURDATE())-Year(d.usu_nacimiento) between '".$edad."' and
'".$edad2."'and a.loc_date >='".$date1."' and a.loc_date
<='".$date2."'" ;

$result = mysqli query($con,$sql);
```

```php
}else{

$sql="select
a.loc_latitude,a.loc_longitude,a.loc_time,a.loc_date,b.seg_activit
y, b.seg_id from tbl_Location a JOIN tbl_Segmento b ON a.seg_id =
b.seg_id JOIN tbl_usuarios d ON b.usu_hash = d.usu_hash where
YEAR(CURDATE())-Year(d.usu_nacimiento) between '".$edad."' and
'".$edad2."' and b.seg_activity='".$actividad."' and a.loc_date
>='".$date1."' and a.loc_date <='".$date2."'" ;

$result = mysqli_query($con,$sql);
}

$path = array();
$tempSeg ="";
$tempAct ="";
$templat ="";
$templon ="";
$tempdate ="";
$temptime ="";
$tempname ="";
$row_cnt = mysqli_num_rows($result);
$cont=0;

while($row = mysqli_fetch_array($result)) {
    //ultimo segmento

    $name=$row['seg_activity'];
    $dateLoc=$row['loc_date'];
    $timeLoc=$row['loc_time'];
    $lat=$row['loc_latitude'];
    $lon=$row['loc_longitude'];
    $segId=$row['seg_id'];
    if ( $cont == 0){
    $tempSeg = $segId;
$tempAct = $name;
$templat = $lat;
$templon = $lon;
$tempdate = $dateLoc;
$temptime = $timeLoc;
$tempname = $name;
    }

    if ($segId == $tempSeg){
     $pair = array($lat,$lon);
     array_push ($path, $pair);
     $tempSeg = $segId;
     $tempAct = $name;
     $templat = $lat;
     $templon = $lon;
     $tempdate = $dateLoc;
     $temptime = $timeLoc;
     $tempname = $name;
```

```php
}else{
        $path = json_encode($path);
        echo ("drawPolylines($path,'$tempAct');\n");
        echo ("addMarker($templat,
$templon,'$tempname','$tempdate','$temptime');\n");
        echo ("addMarker($lat,
$lon,'$name','$dateLoc','$timeLoc');\n");
        $path = array();
        $tempAct ="";
        $tempSeg = "";
        $templat ="";
        $templon ="";
        $tempdate ="";
        $temptime ="";
        $tempname ="";
        $pair = array($lat,$lon);
        array_push ($path, $pair);

    }

    if ($cont == $row_cnt-1 || $cont ==0){
      if ($cont == $row_cnt-1){
      $path = json_encode($path);
      echo ("drawPolylines($path,'$tempAct');\n");
      }
      echo ("addMarker($lat,
$lon,'$name','$dateLoc','$timeLoc');\n");
    }

    $cont++;
$tempSeg = $segId;
$tempAct = $name;
$templat = $lat;
$templon = $lon;
$tempdate = $dateLoc;
$temptime = $timeLoc;
$tempname = $name;

}

mysqli_close($con);

?>
 }
 </script>

<body onload="initMap()" style="margin:0px; border:0px;
padding:0px;">

 <?
 if(mysqli_num_rows($result)==0){
 echo "No se ha encontrado ningun marcador";
 }else{
 if (isset($_POST['submit_map'])) {
        echo "<div id=\"gmap\"\> ";

    }
 }
 ?>
```

```html
<script type="text/javascript" src="/js/jquery.js"></script>
    <script type="text/javascript"
src="/js/bootstrap.min.js"></script>
    <script type="text/javascript"
src="http://maps.google.com/maps/api/js?sensor=true"></script>
    <script type="text/javascript" src="/js/gmaps.js"></script>
    <script type="text/javascript"
src="/js/smoothscroll.js"></script>
    <script type="text/javascript"
src="/js/jquery.parallax.js"></script>
    <script type="text/javascript" src="/js/coundown-
timer.js"></script>
    <script type="text/javascript"
src="/js/jquery.scrollTo.js"></script>
    <script type="text/javascript"
src="/js/jquery.nav.js"></script>
    <script type="text/javascript" src="js/main.js"></script>
</body>
</html>
```

**Chapter 9**